\documentclass[11pt,a4paper]{article}
\usepackage[utf8]{inputenc}
\usepackage[T1]{fontenc}
\usepackage{amsmath,amssymb}
\usepackage{booktabs}
\usepackage{graphicx}
\usepackage{geometry}
\usepackage{natbib}
\usepackage{hyperref}
\usepackage{siunitx}
\usepackage{caption}
\usepackage{placeins}
\hypersetup{colorlinks=true,linkcolor=blue,citecolor=blue,urlcolor=blue,
pdftitle={Measuring Trade Direction in a Prediction Market: Settlement Ground Truth and Trading-Cost Measurement on Polymarket},pdfauthor={Philipp D. Dubach},
pdfkeywords={prediction markets, market microstructure, trade direction, trading costs, blockchain data}}
\graphicspath{{figures/}}

\title{Measuring Trade Direction in a Prediction Market:\\[2pt]
\large Settlement Ground Truth and Trading-Cost Measurement on Polymarket}

\author{Philipp D.~Dubach\\[2pt]
\normalsize\textit{Zurich, Switzerland}\\
\normalsize\href{mailto:phdubach@pm.me}{phdubach@pm.me}}
\date{5 October 2026}

\begin{document}
\maketitle

\begin{abstract}
\noindent
Trade-sign errors can change measured trading costs even when classification accuracy is high.
We validate the side of Polymarket's public trade prints against the taker leg of each print's
on-chain settlement. On twelve selected days between April and August 2026, spanning both
exchange generations, \SI{92.1}{\percent} to \SI{100.0}{\percent} of prints match a settled
taker leg, with exact side and token agreement on all 24.5~million matched pairs. Mint-and-merge
settlement makes that taker leg essential: pooling maker and taker legs changes the measured
buy share. Signing the full cached tape before settlement selection yields 16.6 million prints
signed by every rule; equal-day balanced accuracy is 0.942 for Lee--Ready, 0.768 for the tick
test and 0.670 for retrospective bulk volume classification. On 16.4 million identical eligible
fills, Lee--Ready raises effective spread by 0.429 cents per share under equal-fill weights;
its realised-spread difference is -0.209 cents. Under share-volume weights, taker five-minute
midpoint impact is 0.689 cents, while tick and bulk classifications give -0.374 and -0.119
cents. That aggregate sign reversal disappears when tied print rows are excluded. Distortion
depends on error-weighted signed outcomes, sample selection and weighting. Receipt-time ordering
and unobserved future-quote age limit these delivered-quote accounting quantities; they do not
identify causal impact or private information. Supporting venue and collector analyses provide descriptive diagnostics.
\end{abstract}

\noindent\textbf{Keywords:} prediction markets, market microstructure, trade direction,
trade classification, trading costs, blockchain data

\noindent\textbf{JEL:} G14, G23, D47

\section{Introduction}

Signed spread-based measures of trading costs require knowing which side initiated a trade.
When the side is inferred
from prices and quotes, a small classification error can have a large economic consequence if
errors occur where the trade price lies far from the midpoint. We study this problem on
Polymarket, a prediction market whose orders are matched off-chain and whose trades settle
on-chain. Each public trade print carries a transaction hash. This permits a comparison of its
reported side with the taker order of its own settlement, and permits signed trading-cost
measures to be recomputed against that observed taker side.

Trade-classification studies have long assessed both accuracy and the consequences of errors
\citep{ellis2000accuracy,odderswhite2000occurrence,chakrabarty-2015-bvc,jurkatis-2022-fia}.
\citet{qin-yang-2026-database} also study chain-based direction and downstream proxy bias on
Polymarket V1. A buyer of one outcome can settle against a buyer of its complement through a
mint. Maker and taker legs therefore do not provide interchangeable truth labels, and a pooled
leg benchmark can misstate both the buy share and agreement with the feed. Receipt-time quotes
add a second problem: an accurately reported taker side need not agree with the side inferred
from the last quote delivered to the collector.
These two measurement problems motivate public-feed validation linked to delivered quotes
across both exchange generations, with an explicit complete-set direction convention and
paired accounting of signed-cost distortion.

Our companion study \citep{dubach-2026-anatomy} documented the venue's microstructure across
52 days and 385{,}198 markets using an archive without trade prints. Reconstructing direction
from resting-size decrements recovered the on-chain sign on only about \SI{59}{\percent} of
matched five-second buckets. It left open whether the market channel's
\texttt{last\_trade\_price} event identifies the aggressor. The second archive format studied
here contains that event, and the sample spans the 28 April 2026 migration from CLOB~V1 to
CLOB~V2. The transaction-hash join answers the question without pairing feed and settlement
observations solely by time.

The empirical argument has three parts. First, we validate the unit and coverage of the
reported sign. On twelve selected days between April and August, including two V1-exchange days,
\SI{92.1}{\percent} to \SI{100.0}{\percent} of prints match a settled taker leg under their
transaction hash. Side and token agree on every one of the 24.5~million matched pairs. We also
measure prints without settlements and settlements without prints, so exact agreement on the
matched observations is not confused with completeness of the tape.

Second, we construct signs on every cached print before selecting those with settlement truth,
then compare rules on the same observations. On the 16.6~million prints signed by all four
rules, equal-day balanced accuracy is 0.942 for Lee--Ready, 0.768 for the tick test and 0.670
for the retrospective BVC bar-majority diagnostic. Lee--Ready leads on every sampled date,
but its score ranges from 0.901 to 0.959. Restoring unmatched prints to the history changes
the scores little; rule-specific eligibility changes the comparison more. This distinction
prevents a scoring-population effect from being mistaken for a construction improvement.
A quote strictly before the print is used because message order within a receipt millisecond
is unresolved.

Third, we connect sign errors to economic measurement on identical eligible fills, in cents
of collateral per share. On 16.4 million common observations, Lee--Ready raises effective
spread by 0.429 cents under equal-fill weights, while reducing realised spread by 0.209 cents.
Under share-volume weights, tick and BVC reverse the sign of the average five-minute midpoint
impact relative to the taker; that aggregate reversal disappears when tied scored prints are
removed. The distortion is not determined by accuracy alone: it depends on the error-weighted
signed outcome, and differs with the accounting quantity, weights and eligible population.
The comparison therefore reports baseline levels, paired differences and sensitivity rather
than treating a classification score as proof of an economically reliable measure.

The broader archive supplies context and diagnostics rather than additional main
contributions. Appendix~\ref{app:venue-analysis} retains the migration, tick-change and
quoted-liquidity analyses. The migration moved every market at once and has no contemporaneous
control group; tick changes generally follow prices near the extremes; and the World Cup
comparison contains few independent event groups. Narrower spreads are consistent with a
binding grid, while game progress and the grid's contribution remain unresolved. A
listings-based health rule (Appendix~\ref{app:health}) and the two-way feed--settlement
coverage audit address collector gaps that can otherwise resemble venue events.

The main results concern measurement on twelve observed days in one venue. They do not establish
causal market-quality effects or population accuracy beyond the sampled days.

\section{Related literature}

\paragraph{Prediction markets and calibration.}
The price-level literature establishes that prediction-market prices are informative and
reasonably calibrated when expiry is near, with a longshot bias for events far in the future
\citep{wolfers-zitzewitz-2004,page-clemen-2013}, though a price need not equal the traders' mean
belief \citep{manski-2006}. These price-level questions leave a separate measurement problem:
the cost of taking an informational position depends on the trade side and the quotes used to
measure that cost.

\paragraph{Decentralised prediction-market microstructure.}
\citet{rahman-al-chami-clark-2025} survey the domain and frame its open questions without
bringing matching tick-level data. \citet{tsang-yang-2026-anatomy} study the 2024 US presidential
election markets (chiefly the Trump YES contract) over many time slices and document
episode-level patterns, with $\lambda$ estimated on aggregated rather than tick-level data.
\citet{sirolly-2025-wash} find that a quarter of the venue's lifetime volume looks like wash
trading, a practice that \citet{cong-li-tang-yang-2023-wash} quantify on unregulated
cryptocurrency exchanges. \citet{ng-2025-discovery} compare price discovery across venues, finding that
Polymarket leads around the 2024 presidential election and in politics and entertainment, while
Kalshi leads in sports and when its relative trading activity is higher; within Polymarket,
\citet{gomezcram-2026-skilled} attribute most price discovery to a few percent of accounts.
\citet{xi-moallemi-pai-wang-2026-volatility} model forward-looking volatility in
prediction markets as a function of probability, time to resolution, spreads and activity, and
\citet{dai-jia-yu-2026-settlement} show that a short-duration market settling on an asset price
that its holders can move transfers wealth from liquidity traders to manipulators. On Polymarket's
BTC 15-minute binaries, \citet{young-2026-openmarket} finds that a walk-forward logistic model over
43 microstructure features with Binance order flow does not beat, and slightly underperforms, the
probability already implied by Polymarket's own book, and that simulated trading on it loses a
little. This paper measures the trade sign and makes no forecast with it.

The election-market evidence of \citet{tsang-yang-2026-anatomy} is complemented by the
broader cross-section in our companion study \citep{dubach-2026-anatomy}:
within its 52-day, 385{,}198-market sample, 600 markets were observed simultaneously at tick level,
with microstructure computed on the full event tape joined to the on-chain record.
\citet{qin-yang-2026-database} release the first-generation exchange's on-chain trade record
with aggressor labels and study classification bias in VPIN, trade-based order-flow measures
and forecasting relationships. They deliberately omit off-chain quote flows. Their tick-rule
and BVC raw accuracies, \SI{49.83}{\percent} and \SI{50.51}{\percent}, apply to a filtered
Standard Binary sample spanning the V1 lifecycle, with routing records and specified market
types excluded. Their log-index execution records and YES-axis normalization differ from our
transaction-hash-matched, asset-local taker target. In our settlement-filtered own-rule benchmark on the two
V1 days, tick raw accuracy is 0.704 and 0.710; period, unit, sign convention and sample selection
differ, so these data do not reconcile that gap. Our distinct comparison validates public-feed
prints and pairs their delivered quotes with settlement direction across both exchange
generations to measure signed-cost distortion. We do not claim the first chain-based target
or the first economic-bias analysis on this venue.
\citet{shen-2026-ghost} document settlements that revert on-chain after the venue reported the
trade; we meet the same transactions from the feed's side.

\paragraph{Trade classification.}
In equity data, \citet{ellis2000accuracy} and \citet{odderswhite2000occurrence} find that
the quote-and-tick algorithm of \citet{lee-ready-1991} misclassifies a material share of
trades, concentrated inside the spread and in fast markets.
\citet{chakrabarty2012short} show that it classifies short sales about as accurately as long sales
and that its trade-level accuracy depends on how quotes are timed against trades, and
\citet{chakrabarty-2015-bvc} compare it with bulk volume classification and the tick rule.
\citet{easley2016discerning} argue that bulk classification can be preferable when individual signs
are noisy, and \citet{jurkatis-2022-fia} halves Lee--Ready's error rate at one-second timestamps.
On a centralised cryptocurrency exchange, \citet{ma-zhai-2021-tick} score the tick rule against the
exchange's own side flag at \SI{77}{\percent}. The
order-flow imbalance of \citet{cont-kukanov-stoikov-2014}, which has a strong relationship with
contemporaneous price changes, and the queue imbalance that \citet{gould-bonart-2016} study as a
one-tick-ahead predictor are built from the best quotes and the sizes queued at them and need no
trade sign; the trade imbalance that Cont, Kukanov and Stoikov compare with order-flow imbalance,
Kyle's $\lambda$ and the signed spreads this paper reports do. We use the recorded side of each matched settlement's taker order as the target, and validate
the feed against it by transaction hash. The public chain supplies an auditable target for this
venue, while the mint-and-merge mechanism requires a distinction between takers and the other
legs of the same transaction. Our economic comparison asks how the errors alter the signed
price distances on the same fills, rather than treating a classification score as sufficient
evidence about the reliability of a trading-cost measure. When one side dominates, raw accuracy is a
misleading score and the Matthews correlation a more reliable one \citep{chicco2020advantages}; we
compare the rules on balanced accuracy and report the Matthews correlation beside it. Against the
taker side, the feed's print carries the sign exactly, and Lee--Ready on the quote strictly before
the print recovers it with balanced accuracy of 0.942 on the common scored sample.

\section{Institutional background}

\subsection{Polymarket and the first-generation venue}

Polymarket runs a central limit order book (CLOB) for binary markets: an operator matches orders
off-chain, and trades settle on Polygon in a stablecoin against an on-chain conditional-token
contract. Each market has two tokens, one per outcome, which this paper calls assets. A complete
set, one token of each outcome, redeems for one unit of collateral; it can be minted from
collateral and merged back into it. Until April 2026 the venue operated its first-generation
exchange (``CLOB~V1''): a standard exchange contract and a neg-risk exchange contract, through
which the venue's neg-risk markets settle. Orders were signed under an EIP-712 domain (the Ethereum
standard for signing typed structured data) whose version was \texttt{1} and which embedded a
\texttt{feeRateBps} field in the signed order. The public market channel broadcasts book
snapshots, price-level changes and trade prints. The archive this paper uses kept no prints in its
first format, so in that format the aggressor is not identified, which is the origin of the
measurement problem this paper closes.

\subsection{The CLOB V2 migration}
\label{sec:migration}

On 28 April 2026 the venue shipped a coordinated upgrade: new exchange contracts, a rewritten
matching backend, a new collateral token and a new fee model. Its changelog of 17 April announced
the upgrade for about 11:00 UTC on 28 April, with about an hour of downtime, and told traders to
plan for all open orders to be wiped at the cutover; its entry of 28 April records that resting
orders from before the cutover did not migrate, and the books reopened without them. Two of the changes matter for reading the chain and
the feed.

\paragraph{New exchange contracts.} The V2 exchange addresses differ from V1, and the on-chain
event ABI was rewritten. The fill event gained fields (most importantly an explicit \texttt{uint8}
side), and the \texttt{OrdersMatched} and \texttt{FeeCharged} events, which V1 also emitted,
changed signature. A replication that filters the V2 contracts' logs on the V1 fill event's topic,
the hash of its signature, returns zero legs; Appendix~\ref{app:decoding} gives both signatures.

\paragraph{A new fee model.} V1 embedded the fee in the signed order. V2 collects fees on-chain at
match time, with category-specific taker rates and a fee-free Geopolitics category, and carries
the fee on the fill event (Appendix~\ref{app:decoding}). The feed's \texttt{fee\_rate\_bps} field
is a V1 artefact and is not the V2 fee. This paper does not analyse fee payments; fee rates enter only as
descriptors of the venue's categories in Section~\ref{sec:cutover}.

\section{Data}
\label{sec:data}

\subsection{The feed archive}

The order-book data are a tick-level archive of Polymarket's public WebSocket market channel,
captured by a third party, PMXT \citep{pmxt-archive-v2}, in two formats: the provider's first format from 21 February to 15
April 2026 and its second format from 15 April to 10 August, 171 UTC days in all with the seam day
present in both. The archive's seam precedes the venue's migration, so the last days of the V1
exchange are second-format days. Each event carries a receipt time and a native timestamp, market
and asset identifiers, an event type and, for price-changing messages, the best bid and best ask on
every row. All feed timestamps are the archive's receipt time, not the venue's native time, which
the feed re-sends on some rows of the second format and which is not usable as a partition clock.
The second format adds the trade prints with their transaction hashes, the tick-size change
messages and the book snapshots the feed sends on subscription.

Because best quotes travel on every \texttt{price\_change} row, we recover a per-asset quoted
spread series without replaying the order book. A quote's spread is the ask minus the bid. A
one-sided book, quoted on one side only, arrives with the missing side at 0 or 1, so its spread is
one minus the lone bid or the lone ask itself. A quote counts as empty when its spread is 0.9 or
more, which covers a book with no side and the venue's placeholder (a midpoint of 0.5 with a spread
of 0.9 or more), and as live otherwise; an asset-day is live when its median spread is above zero
and below 0.9.

From both formats we build one panel per (market, asset, UTC day): update counts; the median and
quartiles of the spread over the day's updates, exact to 0.001 (each is the lower edge of the 0.001
bin that holds that quantile's update); spread sums and squares; the last midpoint; the hours in
which the asset was present and the number of updates in which its book was empty; displayed size
at the best bid and ask from the \texttt{price\_change} rows at the best level and from the
snapshots; the first quote, first live quote and first trade of the day; and, for the second-format
days, trade counts and volume. The panel holds 17{,}009{,}350 asset-days on 3{,}081{,}897 assets of
1{,}540{,}951 markets; 93{,}588 assets appear in both formats. Each asset's market comes from the
feed's own market field and its outcome from the venue's metadata, the outcome labels the first
format carries and a registry, in that order, and an outcome is known for a share of 0.9993 of asset-days.

\subsection{On-chain ground truth}
\label{sec:onchain}

Ground truth comes from the events of the standard and neg-risk exchange contracts of each
generation on Polygon, read for twelve whole UTC days: 26 and 27 April on the V1 exchange, 29 and
30 April, the first two full days of the V2 exchange, and eight days stratified over May to August
(13 and 31 May, 5 and 9 June, 1, 7 and 17 July, 5 August), chosen to span the sample, to bracket
the 2 July and 10 July venue changes and to avoid the archive's June and July outages
(Section~\ref{sec:health}). Each day's block range is fixed by block timestamps and extended 300
blocks (seven and a half to ten minutes) past midnight so that the settlements of the day's last
prints are captured. We keep every \texttt{OrderFilled} leg of these contracts, 4.6 to 7.2~million
a day, and every \texttt{OrdersMatched} and \texttt{FeeCharged} event, each with its block
timestamp. The cutover day itself, 28 April, is a boundary and is excluded. Only the V2 fill event
carries a side field; on V1 the side follows from which asset the order's maker gave, so both
exchanges yield a signed taker leg per settlement (Section~\ref{sec:method-match}).

\subsection{Summary statistics}

\begin{table}[htbp]
\centering\footnotesize\setlength{\tabcolsep}{3.5pt}
\begin{tabular}{lrrrrr}
\toprule
& mean & p25 & median & p75 & p95 \\
\midrule
\multicolumn{6}{l}{\textit{Panel A: the panel, 171 days over both archive formats, per asset-day}} \\
Binned median quoted spread, live asset-days & 0.194 & 0.010 & 0.038 & 0.260 & 0.850 \\
Asset-days, live / all & \multicolumn{5}{c}{13{,}451{,}616 / 17{,}009{,}350} \\
Asset-days by format, first / second & \multicolumn{5}{c}{2{,}855{,}452 / 14{,}153{,}898} \\
Updates per asset-day, median & \multicolumn{5}{c}{2{,}163} \\
Assets observed per day, minimum and maximum over clean days & \multicolumn{5}{c}{26{,}928--302{,}256} \\
Markets listed by the venue per day, median February--July / August & \multicolumn{5}{c}{8{,}759 / 23{,}547} \\
\midrule
\multicolumn{6}{l}{\textit{Panel B: the on-chain sample, 12 days}} \\
Feed prints & \multicolumn{5}{c}{25{,}035{,}601} \\
Settled taker legs & \multicolumn{5}{c}{27{,}144{,}192} \\
Prints matched to their taker leg by transaction hash & \multicolumn{5}{c}{24{,}537{,}912} \\
\bottomrule
\end{tabular}
\caption{Summary statistics. Panel A is the harmonised panel of both archive formats; `live' marks live days, asset-days whose binned median spread is above zero and below 0.9; clean days exclude the degraded days of Appendix~\ref{app:health}. Panel B is the on-chain sample of Section~\ref{sec:identity}.}
\label{tab:summary}
\end{table}
 
Table~\ref{tab:summary} summarises the panel and the on-chain sample; Section~\ref{sec:composition}
and Figure~\ref{fig:series} describe how the cross-section moves over the sample.

\subsection{Archive health}
\label{sec:health}

The feed archive is a third party's capture, and its gaps look like market events unless they are
flagged. We benchmark it against the venue's own listings, from a sweep of the venue's metadata API
by listing day, and a rule flags a day as degraded when an hourly file is missing or near-empty,
when the archive first sees almost no new assets while the venue keeps listing markets, or when it
carried fewer than half of the markets listed that day. Appendix~\ref{app:health} gives the rule,
the benchmark and the list. Over the whole sample the archive carried \SI{81.3}{\percent} of the
markets the venue listed (1{,}511{,}091 of 1{,}858{,}106) on the listing day or later; on the
median clean day it carried \SI{99.98}{\percent}, and on the tenth-percentile clean day
\SI{93.6}{\percent}. Thirty-six of the 171 days are degraded, and the rule's first-seen,
carried-share and window thresholds move that count by at most three days
(Appendix~\ref{app:health}).

\section{Methodology}
\label{sec:method}

\subsection{Matching prints to settlements}
\label{sec:method-match}

Every \texttt{last\_trade\_price} event the archive's second format records carries the hash of the
Polygon transaction meant to settle the trade it reports. Each settled transaction on the
V2 exchange contracts emits one \texttt{OrderFilled} leg per order it fills, and exactly one of those
legs has the exchange contract itself in the \texttt{taker} field: the taker order. We join each
print to that leg by transaction hash. On the first-generation exchange the fill event carries no
side, but the side follows from which asset the order's maker gave: collateral for a buy, the
outcome token for a sell. The join is therefore the same on both exchanges. For each matched pair
we compare the side, the token, the price and the size, where the on-chain price and size follow
from the two filled amounts and the side.

The residual runs in both directions. On the feed's side, a print with no settled counterpart is
matched in a second stage to the unmatched taker leg of the same token and side, with the same
price to four decimals and the same size to six, that lies nearest in time within two minutes. An
hour-stratified sample of the prints that remain is classified from the transaction receipt: no
transaction under the hash, reverted, mined outside the scraped block range, or settled by a
contract we did not read; the class shares are weighted by each hour's population, and a sample of
matched hashes is run through the same lookup as a control. On the chain's side, a taker leg with
no print is a fill the archive does not contain; the scrape's tail past midnight is excluded from
that count.

\paragraph{What a settlement looks like.} The exchange settles a taker order in one of three ways,
and the legs show which. A \emph{complementary} match fills the taker against makers on the same
token with the opposite side. A \emph{mint} fills a taker buying one outcome against makers buying
the other, and creates the pair of tokens from collateral; a \emph{merge} is its mirror for sells.
We classify each transaction from its legs; one whose maker legs include both kinds, some on the
taker's token and some on its complement, is \emph{mixed}.

\subsection{Direction rules and their scoring}
\label{sec:method-rules}

The primary benchmark constructs signs on every print in each cached UTC daily tape before
joining to settlement truth. It therefore preserves unmatched prints in the tick history and
five-second bars. Only hash-matched prints can be scored. This separates an implementable
feed-based history from the settlement selection needed for validation; it does not recover
prints missing from the archive. For comparison, a settlement-filtered construction signs only the matched
prints. Both constructions reset at the start of the day.

Lee--Ready signs a print above the prevailing midpoint as a buy, below it as a sell, and at
the midpoint by the tick test. The quote rule signs a print at or above the ask as a buy and
at or below the bid as a sell, using the tick test inside the spread. Missing or crossed quotes
invoke the tick fallback. The tick test compares the price with the asset's previous print and
carries the last non-zero sign through zero ticks; observations before its first non-zero
change are unsigned. Bulk volume classification (BVC) assigns each five-second bar a buy share
from its last-price change scaled by a twenty-bar rolling standard deviation
\citep{easley2016discerning}, then assigns that majority sign to the prints in the bar. An
unchanged or first bar is unsigned. BVC estimates a volume fraction rather than individual
initiators. Its closing price and whole-day fallback volatility are unavailable at each print;
this is a retrospective bar-majority diagnostic, not a streaming classifier. Exact definitions
and the paired history comparison are in Appendix~\ref{app:rules}.

The prevailing quote is the asset's last \texttt{price\_change} strictly before the print's
receipt millisecond. Quotes stamped at that same millisecond are excluded because their order
relative to the print is unresolved. Print histories use the saved cache row order to break
asset--timestamp ties deterministically; that order is not established venue order. A
sensitivity removes tied print rows from scoring while retaining their history. It tests their
influence on the reported scores, but cannot eliminate their possible influence on subsequent
signs. Cached quotes at lags from 1~ms to 1~s permit a further timing comparison.

For each rule we report raw accuracy, balanced accuracy (the mean buy and sell recall), the
Matthews correlation and coverage on its signed subset. The primary ranking additionally uses
the intersection signed by all four rules, so every rule faces identical truth observations
and class composition. The history comparison pairs full- and selected-history predictions on
identical cache row ids, and compares their scores only where both assign a sign. It isolates
construction changes from abstention changes. Across days we report equal-day means, ranges
and observed paired differences. Twelve dates chosen to span venue changes are not random or
necessarily independent; these contrasts describe the sample without a population-significance
claim.

\paragraph{Cell-based aggregation.} Within each asset's five-second cells, we compare feed
direction with all on-chain legs pooled together and with taker legs separately. This
comparison assesses the consequences of pooling settlement legs. Feed cells use receipt times
and settlement cells use block times, so cell disagreement also reflects the clocks.

\subsection{Economic consequences on identical eligible fills}
\label{sec:method-signed}

Let $p_i$ denote the print price, $m_i$ the preceding delivered midpoint, $m'_i$ the cached
midpoint five minutes later and $s_i$ the taker sign. For any sign $a_i$, the accounting quantities
are effective spread $E_i(a)=2a_i(p_i-m_i)$, realised spread $R_i(a)=2a_i(p_i-m'_i)$ and
five-minute midpoint impact $I_i(a)=2a_i(m'_i-m_i)$. Hence $E_i(a)=R_i(a)+I_i(a)$ on every
observation. Results are reported in cents of collateral per share, with both equal-fill and share-volume
weights. They are full-spread quantities; fees, rebates and inventory costs are excluded.

All four rules and the taker are compared on identical observations, using full-tape signs.
Eligibility requires all four signs, a finite print price in probability support, an uncrossed
strictly interior current quote ($0<b_i\leq u_i<1$, with $u_i$ the ask), a finite forward midpoint in probability
support and a positive finite print size. Prints whose five-minute horizon reaches or passes
the exclusive UTC day boundary are excluded. A sequential funnel records each exclusion. The future quote's
timestamp, bid and ask were not saved, so a finite future midpoint does not establish its age,
its two-sided state or an uncrossed future book. The day-end exclusion addresses a known
boundary, not internal gaps or stale quotes. We also restrict current quote age to at most one
second and remove tied print observations as sensitivity checks.

For rule $r$, let $e_i=\mathbf{1}\{r_i\ne s_i\}$. For any of the three measures $M$,
\[
 M_i(r)-M_i(s)=-2e_i M_i(s),\qquad
 \overline M_w(r)-\overline M_w(s)=-2\sum_i w_i e_iM_i(s),
\]
where the non-negative weights sum to one on the common sample. Thus error frequency alone
cannot determine distortion: it is the error-weighted signed outcome that matters. For valid
uncrossed current quotes, off-midpoint Lee--Ready errors are wrong-side fills and increase
effective spread mechanically. No analogous direction of distortion follows for realised
spread or midpoint impact. The code verifies both identities for every sign and weighting.

\begin{quote}\small
\noindent\textbf{Observed signing error.} On 17 July 2026 at 15:36:43.018 UTC, a cached
print reports a buy of 115.07 shares at 0.22. Its own settlement confirms the same buy, token,
price and size. The preceding delivered bid and ask, received 160~ms earlier, are 0.22 and
0.24, giving a midpoint of 0.23. Lee--Ready therefore signs the print as a sell. Effective
spread is $-2$ cents per share under the taker sign and $+2$ cents under Lee--Ready: the
sign error adds 4 cents on this fill. This negative taker-signed distance is relative to the
delivered quote; it does not establish the execution-time venue book or a profitable trade.
Appendix~\ref{app:decoding} identifies the transaction and its settlement legs.
\end{quote}

We report signed differences in price units rather than ratios near a zero denominator, and whether the
sign of a day's mean changes. Exploratory per-asset comparisons use the same eligible assets
with at least fifty fills to describe sign changes and rank correlations; they do not treat
assets sharing a market or event as independent draws. Means with absolute value at most
$10^{-12}$ price units count as numerical zero for sign diagnostics; asset means are rounded to
twelve decimal places only for ranking. These precision conventions leave accounting means
unchanged and are not economic materiality thresholds.

These are accounting comparisons against delivered quotes. Receipt-time ordering does not
identify the execution-time venue book, and a five-minute signed midpoint change is not an
identified causal price impact or measure of private information. Realised spread here is
not a fee-adjusted profitability estimate.
 
\subsection{Inference}

The primary classifier and economic comparisons use the day as the summary unit
(Section~\ref{sec:method-rules}); no uncertainty interval is pooled over millions of fills.
The twelve days were selected to span the sample and venue changes rather than sampled at
random. Their mean scores, ranges and paired contrasts describe these observed days; they do
not provide population inference over all trading days. Inference for the
supporting venue analyses is stated in Appendix~\ref{app:venue-analysis}.
\section{Results}
\label{sec:results}

\subsection{The print is the taker leg}
\label{sec:identity}

\begin{table}[htbp]
\centering\scriptsize\setlength{\tabcolsep}{3pt}
\begin{tabular}{lrrrrrrrr}
\toprule
\multicolumn{9}{l}{\emph{Panel A. Coverage and identity}} \\
day & prints & with leg (\%) & legs in day & with print (\%) & side $\neq$ & token $\neq$ & price (\%) & size (\%) \\
\midrule
26 Apr (V1) & 2{,}383{,}232 & 95.9 & 2{,}399{,}429 & 95.3 & 0 & 0 & 92.3 & 99.96 \\
27 Apr (V1) & 2{,}522{,}118 & 96.8 & 2{,}534{,}187 & 96.4 & 0 & 0 & 92.4 & 99.95 \\
29 Apr & 2{,}169{,}975 & 93.9 & 2{,}150{,}324 & 94.8 & 0 & 0 & 92.5 & 99.93 \\
30 Apr & 2{,}237{,}594 & 92.1 & 2{,}178{,}955 & 94.6 & 0 & 0 & 92.1 & 99.95 \\
13 May & 2{,}251{,}602 & 99.9 & 2{,}271{,}729 & 99.0 & 0 & 0 & 92.0 & 99.96 \\
31 May & 1{,}905{,}025 & 100.0 & 1{,}969{,}244 & 96.6 & 0 & 0 & 91.8 & 99.96 \\
5 Jun & 1{,}705{,}254 & 99.9 & 2{,}349{,}031 & 72.3 & 0 & 0 & 93.0 & 99.97 \\
9 Jun & 2{,}254{,}797 & 99.8 & 2{,}385{,}448 & 94.0 & 0 & 0 & 93.7 & 99.97 \\
1 Jul & 1{,}833{,}608 & 100.0 & 2{,}508{,}976 & 72.6 & 0 & 0 & 95.1 & 99.91 \\
7 Jul & 2{,}242{,}392 & 100.0 & 2{,}547{,}188 & 88.0 & 0 & 0 & 94.4 & 99.90 \\
17 Jul & 1{,}876{,}816 & 100.0 & 1{,}977{,}096 & 94.9 & 0 & 0 & 93.4 & 99.96 \\
5 Aug & 1{,}653{,}188 & 100.0 & 1{,}704{,}195 & 97.0 & 0 & 0 & 92.7 & 99.96 \\
\bottomrule
\end{tabular}
\par\medskip
\begin{tabular}{lrrrrrrr}
\toprule
\multicolumn{8}{l}{\emph{Panel B. Settlement types and buy share}} \\
day & mint (\%) & compl. (\%) & merge (\%) & mixed (\%) & unknown (\%) & buy share, taker & buy share, all legs \\
\midrule
26 Apr (V1) & 67.8 & 21.9 & 3.5 & 6.8 & 0.0 & 0.812 & 0.848 \\
27 Apr (V1) & 67.6 & 21.5 & 3.4 & 7.4 & 0.0 & 0.806 & 0.849 \\
29 Apr & 67.6 & 22.3 & 2.8 & 7.3 & 0.0 & 0.813 & 0.850 \\
30 Apr & 66.9 & 21.8 & 3.3 & 8.1 & 0.0 & 0.807 & 0.847 \\
13 May & 68.8 & 20.8 & 3.2 & 7.2 & 0.0 & 0.819 & 0.852 \\
31 May & 68.6 & 20.3 & 3.7 & 7.4 & 0.0 & 0.825 & 0.854 \\
5 Jun & 72.2 & 19.0 & 3.0 & 5.8 & 0.1 & 0.838 & 0.866 \\
9 Jun & 72.9 & 18.3 & 2.7 & 6.0 & 0.0 & 0.848 & 0.870 \\
1 Jul & 53.5 & 20.5 & 2.6 & 4.2 & 19.2 & 0.832 & 0.861 \\
7 Jul & 62.8 & 21.4 & 2.6 & 5.4 & 7.9 & 0.824 & 0.858 \\
17 Jul & 73.8 & 17.9 & 2.1 & 5.2 & 0.9 & 0.856 & 0.884 \\
5 Aug & 64.9 & 27.1 & 1.8 & 6.0 & 0.1 & 0.811 & 0.847 \\
\bottomrule
\end{tabular}
\par\medskip
\begin{tabular}{lrrrrrrr}
\toprule
\multicolumn{8}{l}{\emph{Panel C. The residual}} \\
day & dup. prints & unmatched prints & 2nd-stage matched & reverted (\%) & no tx (\%) & other (\%) & receipts \\
\midrule
26 Apr (V1) & 1 & 96{,}602 & 63{,}068 & 20.1 & 79.9 & 0.0 & 960 \\
27 Apr (V1) & 2 & 80{,}016 & 45{,}288 & 32.0 & 68.0 & 0.0 & 960 \\
29 Apr & 52 & 132{,}425 & 50{,}502 & 41.1 & 58.9 & 0.0 & 960 \\
30 Apr & 2 & 177{,}137 & 75{,}201 & 21.2 & 78.8 & 0.0 & 960 \\
13 May & 39 & 2{,}070 & 7 & 100.0 & 0.0 & 0.0 & 888 \\
31 May & 1{,}285 & 813 & 158 & 94.2 & 5.8 & 0.0 & 625 \\
5 Jun & 4{,}880 & 1{,}511 & 613 & 87.8 & 12.2 & 0.0 & 674 \\
9 Jun & 9{,}326 & 4{,}247 & 682 & 17.5 & 6.1 & 76.4 & 588 \\
1 Jul & 11{,}641 & 610 & 21 & 99.2 & 0.0 & 0.8 & 491 \\
7 Jul & 0 & 776 & 33 & 99.9 & 0.1 & 0.0 & 577 \\
17 Jul & 0 & 810 & 22 & 100.0 & 0.0 & 0.0 & 669 \\
5 Aug & 0 & 672 & 305 & 73.2 & 26.8 & 0.0 & 329 \\
\bottomrule
\end{tabular}
\caption{Feed prints against on-chain taker legs, joined by transaction hash. Panel~A: the share of prints with a settled taker leg; the taker legs settled inside the day (the scrape's 300-block tail past midnight excluded) and the share of them with a print; the matched pairs on which the side or the token differ; the shares of matched pairs within $5\times10^{-4}$ in price and $10^{-6}$ in size. Panel~B: settlement types as shares of settled transactions, read from the legs' structure; \emph{unknown} is a transaction with a maker leg on a token absent from the market metadata; the buy share over taker legs and over all legs. Panel~C: prints that repeat a hash; the prints the hash join leaves without a settled leg; those matched in a second stage on token, side, price and size within two minutes; and the receipt classes (\emph{reverted}; \emph{no tx}, no transaction under the hash; \emph{other}, on these days always a transaction mined outside the scraped block range) of an hour-stratified sample of the rest, weighted by each hour's population. \emph{receipts} is the number of hashes in the sample.}
\label{tab:fillmatch}
\end{table}
 
On the eight days from May to August, \SI{99.8}{\percent} to \SI{100.0}{\percent} of prints match
a settled taker leg under their own transaction hash; on the four April days, which straddle the
migration, \SI{92.1}{\percent} to \SI{96.8}{\percent} do (Table~\ref{tab:fillmatch}). On the
24.5~million matched pairs the side and the token never differ. The size agrees to $10^{-6}$ on
\SI{99.9}{\percent} of pairs. The price agrees to $5\times10^{-4}$ on \SI{91.8}{\percent} to
\SI{95.1}{\percent} and to half a cent on every pair: the maximum absolute difference is 0.005 on
each of the twelve days, which is what rounding to the cent produces.

Each print is compared with the taker leg of its own transaction; the maker legs are not in the
comparison. The test has power because a print copied from a maker leg would carry the other token
in a mint (\SI{54}{\percent} to \SI{74}{\percent} of settlements) and the opposite side in a
complementary match, and the size would not agree. The agreement is complete within each settlement
type separately. The sample is the matched prints, so on the April days \SI{92}{\percent} to
\SI{97}{\percent} of prints are inside the test. On the two days before the migration the join runs
against the first-generation exchange, whose fill event has no side field, and with the side derived
from the legs the agreement is the same. The feed therefore gives the aggressor's side; it carries
no identities.

\paragraph{Settlements without a print.} Among settlements inside the day, the share with a print
ranges from \SI{72.3}{\percent} to \SI{99.0}{\percent}, and is at least \SI{88}{\percent} on ten of
the twelve days. Part of the gap is the hour after midnight: on 13 May, \SI{22}{\percent} of the
unmatched settlements are in the first hour, before the collector had subscribed to the day's new
markets. The two days at \SI{72}{\percent} to \SI{73}{\percent}, 5 June and 1 July, are days on
which the archive lacked whole markets or delivered late prints. On 1 July, 485{,}746 settlements
involve a maker token the archive never carried, \SI{19.2}{\percent} of the day's transactions; on
5 June the missing settlements are on known markets, but the prints arrive a median 22~seconds after
their block, where on other days they precede it, and 4{,}880 prints repeat a hash. Both look like
collector faults rather than the venue's, although delayed prints and repeated hashes could arise on
either side; Section~\ref{sec:health} and Appendix~\ref{app:health} give the archive's health by
day.

\paragraph{Prints without a settlement.} In April, 80{,}016 to 177{,}137 prints a day have no
settled leg under their hash. Of those, 45{,}288 to 75{,}201 match an unmatched settlement of the
same token, side, price and size within two minutes, and for the remainder a receipt lookup returns
no transaction at all on \SI{59}{\percent} to \SI{80}{\percent} of an hour-weighted sample, a
reverted transaction on the rest. From May on the residual is 610 to 4{,}247 prints a day, reverted
on \SI{73}{\percent} to \SI{100}{\percent} of the sample on every day but 9 June, where
\SI{76}{\percent} were mined outside the day's block range. A control of 100 matched hashes per day
returns a settled transaction on the exchange inside the range for all 100 on each of the ten days
on the second-generation exchange; on the two first-generation exchange days the same control
returns all 100 matched hashes as sent to a contract other than the exchanges, so there the lookup's
other-contract class carries no information. The feed prints a trade when the exchange accepts it
and the chain records whether it settled; the reverted class is the ghost fill of
\citet{shen-2026-ghost}. Whether the April prints without a receipt were resubmitted under another
hash we cannot show from receipts alone. The second-stage match is consistent with resubmission and
is all the evidence we have, but the match carries no false-match rate: on a venue where many trades
are identical in token, side, price and size, some of those two-minute matches will be coincidences,
so the counts are upper bounds on the genuine matches.

\subsection{What direction means on a mint-and-merge exchange}
\label{sec:mintmerge}

By count, the exchange settles most taker orders by minting or merging complete sets rather than
against a resting order on the same token: \SI{53.5}{\percent} to \SI{73.8}{\percent} of settled
transactions are mints, \SI{17.9}{\percent} to \SI{27.1}{\percent} complementary matches,
\SI{1.8}{\percent} to \SI{3.7}{\percent} merges and \SI{4.2}{\percent} to \SI{8.1}{\percent} mixed.
Up to \SI{19.2}{\percent} of transactions on one day cannot be classified because a maker leg is on
a token absent from the market metadata; every such case is a token the archive never carried, and
the share is under \SI{0.1}{\percent} on eight of the twelve days. In a mint every leg is a buy and
in a merge every leg is a sell: the maker legs carry the taker's side on the complement token.
Pooling legs by token therefore counts a mint's makers as buyers of the other outcome and
overstates the buy share. The buy share of taker legs is \SI{80.6}{\percent} to
\SI{85.6}{\percent}; over all legs it is \SI{84.7}{\percent} to \SI{88.4}{\percent}.

Pooling every settlement leg in an asset-by-five-second cell and comparing the pooled sign
with the feed's gives agreement on \SI{86.8}{\percent} to \SI{90.9}{\percent} of cells against
pooled legs, below a majority baseline of 0.906 to 0.947 (its balanced accuracy is 0.741 to
0.802, against a chance level of 0.5); against taker legs it agrees on \SI{93.1}{\percent} to
\SI{97.8}{\percent} of cells, and fill by fill the feed's side agrees on \SI{100}{\percent}. The cells
that still disagree against taker legs are not evidence about the feed's side either: the feed side is
timed by receipt and the chain side by block, and a print and its own settlement often fall in
different cells.

\subsection{Classifier comparisons on the same prints}
\label{sec:rules}

\begin{table}[htbp]
\centering\footnotesize\setlength{\tabcolsep}{3.5pt}
\begin{tabular}{lrrrrr}
\toprule
rule & own BA & own cover (\%) & common BA & common accuracy & common MCC \\
\midrule
Lee--Ready & 0.954 & 100.0 & 0.942 & 0.937 & 0.809 \\
quote rule & 0.953 & 99.8 & 0.941 & 0.932 & 0.799 \\
tick test & 0.735 & 96.0 & 0.768 & 0.708 & 0.409 \\
BVC, 5 s bars & 0.670 & 67.8 & 0.670 & 0.652 & 0.260 \\
\bottomrule
\end{tabular}
\caption{Full cached daily histories, scored against hash-matched taker sides. Cells are equal-day means over twelve selected dates. Own-sample scores condition on each rule assigning a sign; own coverage is the percentage of matched prints signed. Common scores use the same intersection of four assigned signs on each day, 16{,}615{,}817 prints in total, covering 57.7--72.2\% of matched prints by day. BVC is a retrospective five-second bar-majority diagnostic. Means and ranges describe selected days; no population test is implied.}
\label{tab:cached-benchmark}
\end{table}
 
\begin{figure}[htbp]\centering
\includegraphics[width=\linewidth]{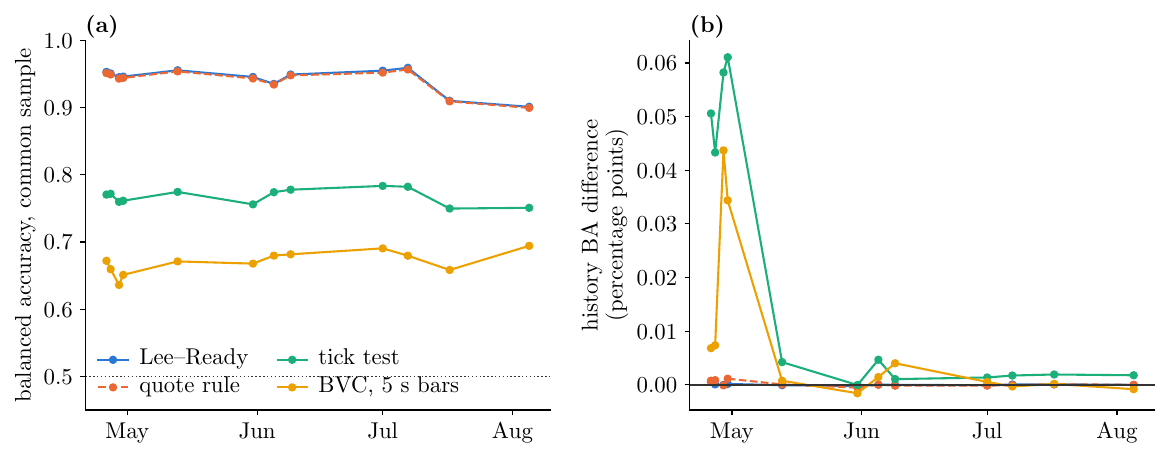}
\caption{Cached-tape comparisons on twelve selected dates. Panel~(a) scores every rule on the
same four-rule signed intersection for each day. Panel~(b) holds the scored rows fixed within
each rule and compares full versus settlement-selected sign histories; the vertical scale is
percentage points of balanced accuracy. Connecting segments guide the eye. BVC is a
retrospective bar-majority diagnostic. No interval or interpolation over unobserved days is
implied.}
\label{fig:cached-direction}
\end{figure}

The full daily caches contain 25.0~million prints, of which 24.5~million match settlement
truth. The intersection signed by all four rules contains 16.6~million prints, covering
\SI{57.7}{\percent} to \SI{72.2}{\percent} of matched prints by day. On this identical
sample, equal-day mean balanced accuracy is 0.942 for Lee--Ready, 0.941 for the quote rule,
0.768 for the tick test and 0.670 for the retrospective BVC majority-bar diagnostic
(Table~\ref{tab:cached-benchmark}). Lee--Ready's day range is 0.901 to 0.959, with the lowest
score on 5 August. Its advantage over the tick test is 0.174 on average (0.150 to 0.190 across
days), and over BVC 0.272 (0.207 to 0.309); both differences are positive on all twelve dates.
The quote rule trails Lee--Ready on each date. These are sample descriptions, not tests over
independent randomly selected trading days.
Figure~\ref{fig:cached-direction} places these daily common-sample scores beside the paired
history contrasts discussed below.

\paragraph{Eligibility matters more than history in this sample.} On each rule's own signed
sample, full-history balanced accuracy is 0.954 for Lee--Ready and 0.735 for the tick test.
The tick test's higher score of 0.768 on the common sample is therefore a scoring-population
change, not an improvement caused by restoring unmatched prints. On identical observations
signed under both histories, the equal-day full-minus-selected change is only 0.000192 for
the tick test and 0.000080 for BVC. The largest daily paired sign-change share is
\SI{0.349}{\percent} for the tick test and \SI{0.377}{\percent} for BVC, and smaller for
the quote-based rules (Table~\ref{tab:cached-history}). Full-tape construction is necessary
for a feed-based benchmark, but its empirical effect here is small. Daily denominators and
truth composition are in Table~\ref{tab:cached-days}.

\paragraph{Timestamp ambiguity and quote timing.} Asset--timestamp print ties affect
\SI{36.9}{\percent} of matched prints. Removing those rows from scoring, while retaining
their history, leaves 9.3~million common observations and raises Lee--Ready's equal-day
balanced accuracy to 0.957. That is a substantial sample change, and cannot establish the
unknown within-tie order or remove its effect on later signs. Keeping one print per matched
hash changes Lee--Ready's common score only to 0.943. On the separate fixed sample where
Lee--Ready and all cached lag variants sign, its balanced accuracy is 0.9538 with the preceding
quote, 0.953 with a 100~ms lag, 0.921 with a 500~ms lag and 0.885 with a one-second lag.
Those figures have a different denominator from the primary four-rule intersection. Quote
lags reveal sensitivity to delivered-quote timing; they do not recover the execution-time
venue book.

\subsection{Trading-cost distortion and the five-minute decomposition}
\label{sec:economic-results}

The accounting sample contains 16{,}448{,}261 prints signed by every rule with valid current
quotes, finite outcomes and sizes, and a five-minute horizon strictly inside the cached day.
It covers \SI{67.03}{\percent} of settlement matches; daily exclusions and fresh-quote counts
appear in Table~\ref{tab:cached-economic-eligibility}. This further selected sample is shared
by all signs and both weightings. Each reported summary averages twelve daily means equally.
Table~\ref{tab:cached-economic-means} gives the levels, and
Table~\ref{tab:cached-economic-distortion} gives paired rule-minus-taker differences.

\begin{table}[htbp]
\centering\footnotesize
\begin{tabular}{lrrr}
\toprule
\multicolumn{4}{l}{\emph{Equal-fill weights}} \\
sign & effective & realised & 5-min impact \\
\midrule
taker & 2.328 & 6.486 & $-4.158$ \\
Lee--Ready & 2.757 & 6.277 & $-3.520$ \\
quote rule & 2.662 & 6.169 & $-3.507$ \\
tick test & 1.475 & 5.279 & $-3.804$ \\
BVC, 5 s bars & 1.204 & 4.692 & $-3.488$ \\
\bottomrule
\end{tabular}
\par\medskip
\begin{tabular}{lrrr}
\toprule
\multicolumn{4}{l}{\emph{Share-volume weights}} \\
sign & effective & realised & 5-min impact \\
\midrule
taker & 2.033 & 1.344 & 0.689 \\
Lee--Ready & 2.432 & 1.708 & 0.724 \\
quote rule & 2.340 & 1.642 & 0.698 \\
tick test & 1.545 & 1.919 & $-0.374$ \\
BVC, 5 s bars & 1.208 & 1.327 & $-0.119$ \\
\bottomrule
\end{tabular}
\caption{Signed accounting quantities in cents of collateral per share, on the identical primary eligible fills for every rule and the taker. Cells average the twelve daily means equally. Each daily equal-fill or share-volume weighting is shared across all signs. Effective equals realised plus five-minute midpoint impact before rounding. The quote is delivered to the collector; these are not fee-adjusted profits or identified causal impact/information estimates.}
\label{tab:cached-economic-means}
\end{table}
\begin{table}[htbp]
\centering\footnotesize
\begin{tabular}{lrrrrrr}
\toprule
\multicolumn{7}{l}{\emph{Equal-fill weights}} \\
rule & $\Delta E$ & $\Delta R$ & $\Delta I$ & flip E & flip R & flip I \\
\midrule
Lee--Ready & 0.429 & $-0.209$ & 0.638 & 0 & 0 & 0 \\
quote rule & 0.334 & $-0.317$ & 0.650 & 0 & 0 & 0 \\
tick test & $-0.853$ & $-1.207$ & 0.353 & 0 & 0 & 0 \\
BVC, 5 s bars & $-1.124$ & $-1.793$ & 0.669 & 0 & 0 & 0 \\
\bottomrule
\end{tabular}
\par\medskip
\begin{tabular}{lrrrrrr}
\toprule
\multicolumn{7}{l}{\emph{Share-volume weights}} \\
rule & $\Delta E$ & $\Delta R$ & $\Delta I$ & flip E & flip R & flip I \\
\midrule
Lee--Ready & 0.399 & 0.365 & 0.035 & 0 & 0 & 0 \\
quote rule & 0.307 & 0.298 & 0.009 & 0 & 0 & 0 \\
tick test & $-0.487$ & 0.575 & $-1.063$ & 0 & 0 & 8 \\
BVC, 5 s bars & $-0.824$ & $-0.016$ & $-0.808$ & 0 & 0 & 6 \\
\bottomrule
\end{tabular}
\caption{Rule minus taker signed means in cents of collateral per share, with the same rows and weights; equal-day means of paired differences. E, R and I denote effective spread, realised spread and five-minute midpoint impact. Flip columns count dates with opposite daily mean signs (out of twelve); absolute means at or below $10^{-12}$ price units are treated as numerical zero. Differences preserve direction rather than taking absolute values. They measure distortion relative to delivered-quote accounting and do not identify changes in market behavior.}
\label{tab:cached-economic-distortion}
\end{table}
 
Under equal-fill weights, the taker effective spread is 2.328 cents of collateral per share,
realised spread is 6.486 cents and five-minute midpoint impact is $-4.158$ cents. Realised
spread exceeds effective spread because the taker-signed delivered midpoint subsequently moves
in the opposite direction on average. Lee--Ready adds 0.429 cents to effective spread, removes
0.209 cents from realised spread and adds 0.638 cents to midpoint impact. Its effective-spread
increase occurs on every sampled date; the realised-spread difference is negative on eleven.
All rules and the taker retain the same daily mean signs under this weighting. A high score
therefore coexists with a cost distortion, without a daily aggregate sign reversal.

Weighting changes the economic result. With printed-share-volume weights, taker effective
spread is 2.033 cents, realised spread is 1.344 cents and midpoint impact is positive at
0.689 cents. Lee--Ready adds 0.399 cents to effective spread and 0.365 cents to realised
spread, leaving impact close to the taker at 0.724 cents. Tick and BVC instead give negative
mean impact, $-0.374$ and $-0.119$ cents, with paired differences of $-1.063$ and $-0.808$
cents. Their daily impact means have opposite signs to the taker on eight and six dates,
respectively; Lee--Ready and the quote rule have none. This is a statement about a delivered-quote
accounting measure. Neither its positive taker value nor a classifier's negative value identifies
informed trading or causal price impact.

\begin{figure}[htbp]
\centering
\includegraphics[width=\textwidth]{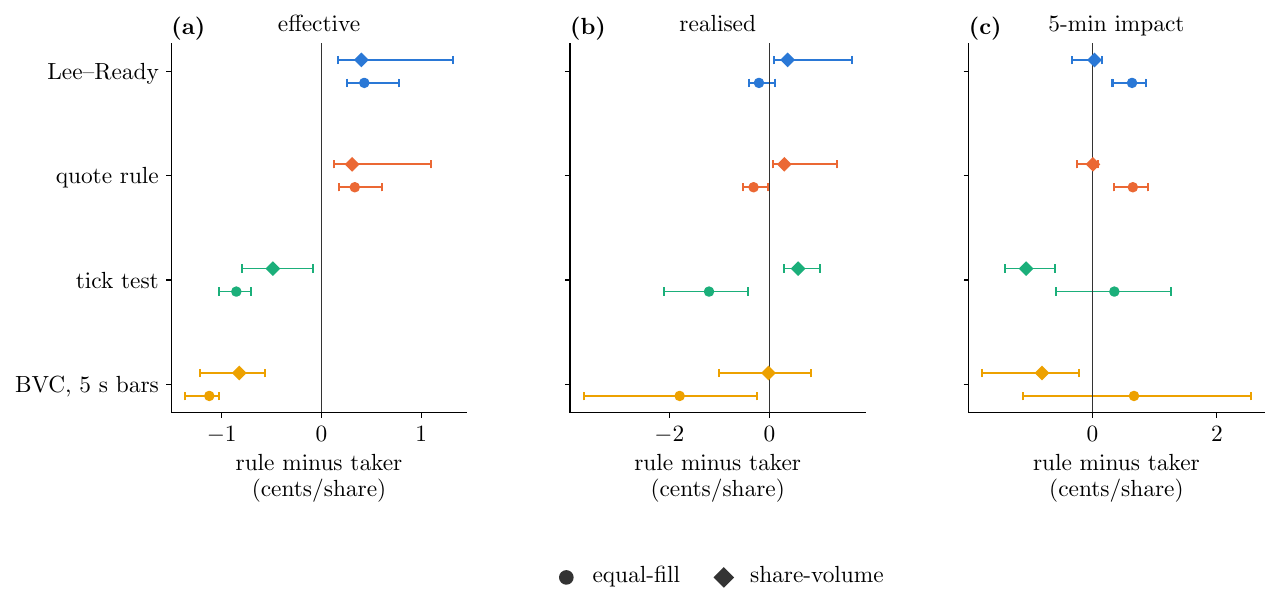}
\caption{Paired rule-minus-taker distortions in cents of collateral per share on identical primary eligible fills.
Markers are equal-day means; circles use equal-fill weights and diamonds use printed-share-volume weights.
Whiskers span the observed daily minimum and maximum, not confidence intervals. Signing rules and
weights change the three accounting quantities differently; the twelve selected dates do not
support a population-significance interpretation.}
\label{fig:cached-economics}
\end{figure}

The effect is outcome-dependent, rather than a universal attenuation factor
(Figure~\ref{fig:cached-economics}). Lee--Ready's equal-fill effective-spread distortion is
entirely upward; its error-weighted signed effective distance is negative. For realised spread,
positive and negative per-fill differences mostly cancel: the average absolute paired distance
is 6.786 cents although the signed difference between means is only $-0.209$ cents. Accuracy
and absolute per-fill error cannot replace the signed error-mass identity in
Section~\ref{sec:method-signed}. In particular, the effective-spread result does not imply a
similar realised-spread or midpoint-impact distortion under another weighting.

\paragraph{Sensitivity and asset comparisons.}
Table~\ref{tab:cached-economic-sensitivity} reports the volume-weighted impact comparison.
Restricting the current quote to an age of at most one second retains negative tick/BVC
aggregate impact and gives nine and seven daily opposite signs, but Lee--Ready then reverses
one daily mean. One eligible row per hash preserves the primary eight/six counts. Substituting
the on-chain price leaves impact unchanged by construction and changes Lee--Ready's equal-fill
effective distortion only from 0.429 to 0.425 cents.

Removing tied scored prints retains 9.1 million observations and changes the aggregate finding:
taker, tick and BVC volume-weighted impact are all positive, at 0.955, 0.102 and 0.181 cents.
Daily tick/BVC reversals fall to five and two. Thus the primary aggregate sign reversal is
sensitive to this selection. This check retains tied observations in the signing history;
it neither recovers venue order nor establishes which scoring population is preferable.

Exploratory ranks use 49{,}817 asset-day observations with at least fifty common fills.
Table~\ref{tab:cached-economic-ranks} reports daily rank correlations and opposite-sign shares.
Under equal-fill weights, Lee--Ready's realised-spread rank correlation with the taker averages
0.982, compared with 0.612 for tick and 0.519 for BVC. Their asset-level realised-spread means
have opposite signs in \SI{3.31}{\percent}, \SI{21.73}{\percent} and \SI{24.61}{\percent}
of the eligible assets on an average day. The primary absence of aggregate daily sign reversals
therefore does not establish preservation of every asset's economic ordering. These dependent,
selected asset comparisons are descriptive rather than tests on independent market draws.
 
\section{Discussion}

The exact side and token agreement in Table~\ref{tab:fillmatch} establishes what the feed reports on matched settlements:
the print corresponds to the settling taker order. Researchers retaining this side field can
construct signed measures directly. An archive without trade prices cannot implement the
price-based rules benchmarked here, and an archive without the side field must infer direction.
The distinction is operational: validation of an existing field and reconstruction of a missing
one solve different data problems.

The effective-spread comparison shows why accuracy is an incomplete criterion for choosing a
trade-classification rule. A wrong sign changes $2s(p-m)$ by $-4s(p-m)$, so errors on fills with
large price distances have larger consequences. For valid uncrossed quotes, Lee--Ready turns
off-midpoint negative taker-signed distances into positive ones by construction. The resulting
upward distortion need not resemble the attenuation produced by independent sign noise.
The paired differences in Table~\ref{tab:cached-economic-distortion} measure this distortion
on the same fills as the taker baseline. Its magnitude belongs to this venue's quotes, settlement mechanism and receipt clock;
classification performance on another venue need not imply the same economic distortion.

The binary settlement mechanism also matters for comparisons with previous evidence. Most
settlements in this sample are mints, in which the maker and taker buy opposite outcome tokens.
Counting every leg as a separate direction observation changes the question being asked.
Likewise, scoring each rule only when it assigns a sign changes the eligible population, and
constructing histories after a settlement filter changes the inputs to tick-based rules.
Figure~\ref{fig:cached-direction} separates the common-sample ranking from the paired
history comparison. Actual history effects are small on these cached days, while the common
sample changes the tick test's balanced accuracy materially.
The primary ranking therefore uses one population, while separate own-rule coverage and
paired history comparisons distinguish eligibility from construction.

Receipt-time ordering sets a limit on the economic interpretation. A taker buy below the last
delivered midpoint need not mean that the taker received a favorable execution against the
book at the venue. The quote may have been delivered out of sequence relative to the print;
same-millisecond ties cannot be ordered from these data. The lag sensitivity is informative
about this uncertainty but does not recover the venue clock. Signed costs computed here are
relative to observed delivered quotes. The negative equal-fill and positive share-volume
taker midpoint-impact averages describe different weightings of this tape. Forward quote age
and future bid/ask are unobserved; a finite cached midpoint and a within-day horizon do not
validate the future book. These responses identify neither private information nor causal
price impact. Removing tied scored prints changes the primary aggregate tick/BVC sign reversal,
so that conclusion also depends on the observed scoring population.

The supporting liquidity analyses show further limits to comparisons on this archive
(Appendix~\ref{app:venue-analysis}). The migration has no unaffected market control, and
displayed size is not comparable across it. Ordinary tick changes depend on extreme prices
near resolution, so their event studies describe the lifecycle. The World Cup arms contain
only six and three treated event groups; their descriptive narrowing cannot separate game
progress from the tick. Finally, a health rule based on new listings cannot detect every loss
of older markets that are still trading. Feed--settlement coverage is therefore a separate
check on completeness rather than a substitute for the listings benchmark.

The twelve on-chain days were chosen to span the archive and venue changes. They establish
exact sign agreement within the matched sample and day-specific coverage, with substantial
variation in classifier performance and signed-cost distortion. They are not a probability
sample of trading days. Dependence among contracts of the same market or event and the small
number of days limit inference beyond the observed sample. The results should be read as a
replicable measurement study of one venue and collector, with explicit boundaries on what the
truth labels and quotes observe.

\section{Conclusion}

Polymarket's public trade prints report the settling taker order's side and token on all
24.5~million matched pairs in twelve selected days. Transaction-hash matching and a separate
audit of missing prints and settlements make that result both precise and conditional.
Mint-and-merge settlement explains why pooled chain legs are an inappropriate substitute for
the taker target.

The full-tape common benchmark gives Lee--Ready balanced accuracy of 0.942, but accuracy does
not determine signed-cost distortion. On identical eligible fills, it adds 0.429 cents per
share to effective spread under equal-fill weights. Its realised-spread effect differs in sign
between equal-fill and share-volume weights. Tick and BVC can reverse the sign of volume-weighted
five-minute midpoint impact relative to the taker, although this aggregate reversal disappears
when tied scored rows are excluded. The relevant mechanism is the error-weighted signed price
distance, rather than error frequency alone. Validated signs, common populations, weights and
quote timing are necessary to interpret these accounting quantities. The supporting venue
and archive analyses retain descriptive evidence while leaving their causal roles unresolved.

\clearpage
\section*{Declarations}

\paragraph{Data availability.} The on-chain events (\texttt{OrderFilled}, \texttt{OrdersMatched}
and \texttt{FeeCharged} of the second-generation exchange contracts and their first-generation
counterparts) are public on Polygon and are re-read by the accompanying scraper from any archive
node; the decoded ABIs and the block ranges of the twelve days are versioned. The order-book feed
is a third-party capture of Polymarket's public WebSocket market channel (PMXT, \url{https://archive.pmxt.dev})
in its first and second formats, 1.8~TB of hourly Parquet files with the provider's receipt files;
it is not redistributed here. The derived inputs (the asset-day panel, the tick-change contexts,
the feed prints and on-chain events, and the auxiliary tick-event panel) are publicly deposited with
an archived code snapshot in the replication package \citep{dubach-2026-replication}. The manifest hashes the
derived files and reuses the downloader's SHA-256 receipts for the archive's second-format
hours. First-format hours have no receipts and are recorded by filename, size and modification
time rather than a content hash. Market metadata (categories,
fee schedules, tick sizes, listing and resolution dates) is a snapshot of the venue's public Gamma
API taken on 26 September 2026 and included as raw pages with a manifest.

\paragraph{Code availability.} The analysis is a set of scripts in the repository directory
\texttt{docs/\allowbreak paper-b/\allowbreak v2-revision/}, one per result, each writing a JSON
file from which the tables and figures are generated. A claims map links numeric statements
to results or declared external facts and design choices. The manifest identifies the inputs
and hashes the derived files, result JSONs and scripts; the environment is pinned. Development
provenance and archived specifications are documented separately in the replication notes.
The analysis code, result files and derived inputs are archived in Zenodo version~1.4
at \url{https://doi.org/10.5281/zenodo.23164515}. This release includes the full-tape
common-sample and three-measure analyses reported here, together with the reproduction of
the observed print illustrated in Section~\ref{sec:method-signed}.

\paragraph{Use of automated assistance.} Large language models were used as coding and writing
assistants for the analysis code, the text, the figures and reviews of drafts.

\paragraph{Funding and competing interests.} The author received no funding for this work and
declares no competing interests.
\clearpage
\bibliographystyle{plainnat}
\bibliography{refs}

\clearpage
\appendix
\section{Direction rules as implemented}
\label{app:rules}

The rules sign prints using feed prices and quotes without the side field. These definitions
apply to both the full-tape and settlement-filtered constructions. The full-tape
history retains all cached prints before scoring hash-matched settlements; the settlement-filtered history
retains only matched prints before signing. All four use the archive's second format.

\paragraph{Prevailing quote.} For each print the prevailing quote is the last \texttt{price\_change}
of the same asset with a receipt time strictly before the print's; a quote stamped at the print's
millisecond may include a post-trade book and is excluded because the order of tied messages
is unresolved. Quotes are attached hour by hour:
the hour's \texttt{price\_change} rows are read for the assets that printed in that hour, joined
as-of in memory, and every asset's last quote of the hour is carried into the next hour, so a print
in the first seconds of an hour sees the quote that preceded it. The midpoint five minutes after a
print is attached the same way, with prints whose horizon crosses the hour boundary waiting one
hour, and so is the last quote before each of five lags behind the print (1, 10, 100, 500 and
1000~ms), which the sensitivity analysis uses.

\paragraph{Lee--Ready.} A print above the prevailing midpoint is a buy, below it a sell, and at the
midpoint the tick test decides \citep{lee-ready-1991}. A print with a missing or crossed prevailing quote also falls
to the tick test.

\paragraph{Quote rule.} A print at or above the prevailing ask is a buy, at or below the bid a
sell, and inside the spread the tick test decides. This is the rule of \citet{ellis2000accuracy},
except that a print beyond a quote is signed by that quote and not by the tick test.
Missing or crossed prevailing quotes invoke the tick fallback.

\paragraph{Tick test.} A print above the asset's previous print in its construction is a buy, below it a sell,
and a zero tick takes the last non-zero sign of the asset. An asset's prints before its first
non-zero tick of the day have no sign; the rule's coverage is reported with its accuracy.

\paragraph{Bulk volume classification.} The construction's prints are grouped into five-second bars per
asset. Each bar's buy share is $\Phi(\Delta p/\sigma)$, with $\Phi$ the standard normal
distribution function, $\Delta p$ the change in the bar's last price from the asset's previous bar
and $\sigma$ the standard deviation of $\Delta p$ over a window of twenty bars ending at the
current one, computed once three bar-to-bar changes exist and otherwise the standard deviation of
$\Delta p$ over all assets and bars of the day \citep{easley2016discerning}. A bar with buy share
above one half is signed as a buy and below one half as a sell. A bar with no price change and an
asset's first bar of the day are unsigned, so coverage depends on the occurrence of price changes. Every print in a bar takes the bar's sign. Bar closing prices
and the whole-day fallback volatility make this diagnostic retrospective.

\paragraph{Scoring.} A print's truth is the side of the taker leg it matches. Recall on each side is
the share correctly signed among that side's prints to which the rule assigns a sign; balanced accuracy is their mean;
$\kappa$ corrects raw agreement for the agreement two independent classifiers with the same
marginals would reach; MCC is the correlation of the two binary labels. Comparisons use
identical scored observations and descriptive equal-day summaries.

\FloatBarrier
\subsection{Sign histories and daily scoring samples}
\label{app:cached-benchmark}

Table~\ref{tab:cached-history} compares the print histories used to construct each rule.
Table~\ref{tab:cached-days} reports the daily scoring samples and agreement metrics.

\begin{table}[htbp]
\centering\footnotesize\setlength{\tabcolsep}{3pt}
\begin{tabular}{lrrrrrr}
\toprule
rule & full BA & selected BA & $\Delta$ BA & range $\Delta$ & flipped (\%) & range (\%) \\
\midrule
Lee--Ready & 0.954 & 0.954 & 0.0000 & [0.0000, 0.0000] & 0.001 & [0.000, 0.003] \\
quote rule & 0.953 & 0.953 & 0.0000 & [0.0000, 0.0000] & 0.005 & [0.000, 0.019] \\
tick test & 0.735 & 0.734 & 0.0002 & [0.0000, 0.0006] & 0.089 & [0.004, 0.349] \\
BVC, 5 s bars & 0.670 & 0.670 & 0.0001 & [0.0000, 0.0004] & 0.097 & [0.001, 0.377] \\
\bottomrule
\end{tabular}
\caption{History construction comparison on identical cache row ids where both histories assign the rule a sign. Full retains all cached prints before matching; selected retains only settlement-matched prints before signing. BA denotes balanced accuracy; $\Delta$ is full minus selected. Score columns and flip share are equal-day means; brackets give day ranges. The paired sample is rule-specific and differs from the four-rule common sample. Signs reset each UTC day. Changes in abstention are recorded separately in the result JSON.}
\label{tab:cached-history}
\end{table}
\begin{table}[htbp]
\centering\scriptsize\setlength{\tabcolsep}{3.5pt}
\begin{tabular}{lrrrrrrr}
\toprule
day & matched & common & cover (\%) & buys (\%) & LR BA & tick BA & BVC BA \\
\midrule
26 Apr (V1) & 2{,}286{,}630 & 1{,}494{,}834 & 65.4 & 82.3 & 0.953 & 0.771 & 0.672 \\
27 Apr (V1) & 2{,}442{,}102 & 1{,}714{,}426 & 70.2 & 82.3 & 0.951 & 0.772 & 0.660 \\
29 Apr & 2{,}037{,}550 & 1{,}439{,}518 & 70.6 & 82.3 & 0.945 & 0.760 & 0.636 \\
30 Apr & 2{,}060{,}457 & 1{,}448{,}512 & 70.3 & 81.6 & 0.946 & 0.762 & 0.651 \\
13 May & 2{,}249{,}532 & 1{,}488{,}463 & 66.2 & 83.2 & 0.956 & 0.775 & 0.671 \\
31 May & 1{,}904{,}212 & 1{,}279{,}945 & 67.2 & 84.0 & 0.946 & 0.756 & 0.668 \\
5 Jun & 1{,}703{,}743 & 1{,}209{,}757 & 71.0 & 84.3 & 0.935 & 0.774 & 0.680 \\
9 Jun & 2{,}250{,}550 & 1{,}591{,}138 & 70.7 & 85.7 & 0.949 & 0.778 & 0.682 \\
1 Jul & 1{,}832{,}998 & 1{,}057{,}960 & 57.7 & 81.2 & 0.955 & 0.784 & 0.691 \\
7 Jul & 2{,}241{,}616 & 1{,}410{,}618 & 62.9 & 82.0 & 0.959 & 0.782 & 0.680 \\
17 Jul & 1{,}876{,}006 & 1{,}354{,}666 & 72.2 & 86.0 & 0.910 & 0.750 & 0.658 \\
5 Aug & 1{,}652{,}516 & 1{,}125{,}980 & 68.1 & 78.7 & 0.901 & 0.751 & 0.694 \\
\bottomrule
\end{tabular}
\caption{Daily denominators and balanced accuracy on the full-history common sample. Buy share is computed on that same sample. V1 denotes the two first-generation exchange dates. The common intersection is a selected subset of matched prints, not representative of all venue transactions. All four rules share these observations; the quote rule is tabulated in the JSON.}
\label{tab:cached-days}
\end{table}
 
\FloatBarrier
\subsection{Economic sample, sensitivity and asset comparisons}

The following tables report economic-sample eligibility, alternative quote and timing
screens, and exploratory comparisons of asset-level means. Each comparison uses identical
eligible fills under the rule and taker signs.
\begin{table}[htbp]
\centering\scriptsize\setlength{\tabcolsep}{3pt}
\begin{tabular}{lrrrrrrr}
\toprule
day & matched & signed & interior & primary & cover (\%) & fresh & assets \\
\midrule
26 Apr (V1) & 2{,}286{,}630 & 1{,}494{,}834 & 1{,}485{,}761 & 1{,}479{,}616 & 64.7 & 1{,}341{,}848 & 4{,}599 \\
27 Apr (V1) & 2{,}442{,}102 & 1{,}714{,}426 & 1{,}704{,}375 & 1{,}700{,}629 & 69.6 & 1{,}567{,}099 & 4{,}613 \\
29 Apr & 2{,}037{,}550 & 1{,}439{,}518 & 1{,}427{,}355 & 1{,}423{,}927 & 69.9 & 1{,}308{,}652 & 3{,}811 \\
30 Apr & 2{,}060{,}457 & 1{,}448{,}512 & 1{,}436{,}746 & 1{,}432{,}080 & 69.5 & 1{,}322{,}697 & 4{,}008 \\
13 May & 2{,}249{,}532 & 1{,}488{,}463 & 1{,}477{,}628 & 1{,}473{,}601 & 65.5 & 1{,}351{,}562 & 4{,}492 \\
31 May & 1{,}904{,}212 & 1{,}279{,}945 & 1{,}270{,}108 & 1{,}266{,}678 & 66.5 & 1{,}169{,}482 & 4{,}047 \\
5 Jun & 1{,}703{,}743 & 1{,}209{,}757 & 1{,}202{,}300 & 1{,}201{,}179 & 70.5 & 1{,}100{,}841 & 3{,}848 \\
9 Jun & 2{,}250{,}550 & 1{,}591{,}138 & 1{,}581{,}911 & 1{,}576{,}556 & 70.1 & 1{,}492{,}041 & 4{,}586 \\
1 Jul & 1{,}832{,}998 & 1{,}057{,}960 & 1{,}050{,}135 & 1{,}046{,}111 & 57.1 & 953{,}170 & 3{,}226 \\
7 Jul & 2{,}241{,}616 & 1{,}410{,}618 & 1{,}400{,}783 & 1{,}396{,}790 & 62.3 & 1{,}305{,}779 & 4{,}218 \\
17 Jul & 1{,}876{,}006 & 1{,}354{,}666 & 1{,}344{,}446 & 1{,}339{,}147 & 71.4 & 1{,}310{,}274 & 4{,}149 \\
5 Aug & 1{,}652{,}516 & 1{,}125{,}980 & 1{,}115{,}943 & 1{,}111{,}947 & 67.3 & 1{,}084{,}822 & 4{,}220 \\
\bottomrule
\end{tabular}
\caption{Sequential economic eligibility: settlement-matched rows; all-four signed rows; rows remaining through current interior uncrossed quotes (after valid print-price screening); final primary rows after forward-midpoint, positive-size and exclusive day-horizon screening. Fresh rows additionally require a current quote no more than one second old. Assets have at least fifty common eligible fills on that date; counts are asset-days and do not imply independence. The future quote age and book state are not stored in the cache.}
\label{tab:cached-economic-eligibility}
\end{table}
\begin{table}[htbp]
\centering\scriptsize\setlength{\tabcolsep}{3pt}
\begin{tabular}{lrrrrrrrr}
\toprule
sample & fills & taker & LR & tick & BVC & flip LR & flip tick & flip BVC \\
\midrule
primary & 16{,}448{,}261 & 0.689 & 0.724 & $-0.374$ & $-0.119$ & 0 & 8 & 6 \\
current quote age $\leq$ 1 s & 15{,}308{,}267 & 0.622 & 0.669 & $-0.707$ & $-0.401$ & 1 & 9 & 7 \\
untied scored rows & 9{,}142{,}989 & 0.955 & 0.930 & 0.102 & 0.181 & 0 & 5 & 2 \\
one eligible row/hash & 16{,}434{,}114 & 0.690 & 0.726 & $-0.372$ & $-0.119$ & 0 & 8 & 6 \\
on-chain price & 16{,}448{,}261 & 0.689 & 0.724 & $-0.374$ & $-0.119$ & 0 & 8 & 6 \\
\bottomrule
\end{tabular}
\caption{Five-minute midpoint impact under printed-share-volume weights, in cents per share; cells average daily means equally. Opposite-sign columns count daily rule/taker mean reversals out of twelve dates, with the stated numerical-zero tolerance. Each row compares all signs on identical observations, but populations differ across sensitivity rows. Signing histories are retained. The chain-price row holds signs, weights and midpoints fixed, so midpoint impact is invariant to its price substitution. Untied scoring removes the aggregate tick/BVC sign reversal; it does not establish venue ordering.}
\label{tab:cached-economic-sensitivity}
\end{table}
\begin{table}[htbp]
\centering\footnotesize
\begin{tabular}{lrrrrrr}
\toprule
\multicolumn{7}{l}{\emph{Equal-fill weights}} \\
rule & $\rho_E$ & $\rho_R$ & $\rho_I$ & flip E (\%) & flip R (\%) & flip I (\%) \\
\midrule
Lee--Ready & 0.935 & 0.982 & 0.983 & 0.1 & 3.3 & 3.1 \\
quote rule & 0.932 & 0.980 & 0.981 & 0.1 & 3.6 & 3.4 \\
tick test & 0.801 & 0.612 & 0.618 & 1.2 & 21.7 & 23.3 \\
BVC, 5 s bars & 0.726 & 0.519 & 0.530 & 4.5 & 24.6 & 25.5 \\
\bottomrule
\end{tabular}
\par\medskip
\begin{tabular}{lrrrrrr}
\toprule
\multicolumn{7}{l}{\emph{Share-volume weights}} \\
rule & $\rho_E$ & $\rho_R$ & $\rho_I$ & flip E (\%) & flip R (\%) & flip I (\%) \\
\midrule
Lee--Ready & 0.905 & 0.949 & 0.951 & 1.1 & 5.7 & 5.2 \\
quote rule & 0.906 & 0.947 & 0.950 & 1.0 & 5.8 & 5.2 \\
tick test & 0.764 & 0.671 & 0.678 & 4.4 & 20.1 & 20.0 \\
BVC, 5 s bars & 0.638 & 0.562 & 0.573 & 10.0 & 22.9 & 22.8 \\
\bottomrule
\end{tabular}
\caption{Exploratory per-asset daily mean comparisons on identical eligible fills and assets with at least fifty observations. The $\rho$ columns correlate rule-signed and taker-signed asset means: E is effective spread, R realised spread, and I five-minute midpoint impact. Rank correlations and opposite-sign asset shares are averaged equally across the twelve dates. Spearman inputs are rounded to twelve decimals to stabilize numerical ties; negligible means use the same $10^{-12}$ zero convention. Assets can share markets/events; no independent-asset inference or population claim is made. These are numerical zero conventions, not economic materiality cutoffs.}
\label{tab:cached-economic-ranks}
\end{table}

\FloatBarrier
\section{On-chain decoding and settlement types}
\label{app:decoding}

\paragraph{Events.} We read the events of four exchange contracts on Polygon:
\begin{center}\footnotesize
\begin{tabular}{ll}
V2 standard exchange & \texttt{0xe111180000d2663c0091e4f400237545b87b996b} \\
V2 neg-risk exchange & \texttt{0xe2222d279d744050d28e00520010520000310f59} \\
V1 standard exchange & \texttt{0x4bfb41d5b3570defd03c39a9a4d8de6bd8b8982e} \\
V1 neg-risk exchange & \texttt{0xc5d563a36ae78145c45a50134d48a1215220f80a} \\
\end{tabular}
\end{center}
On the second-generation contracts we read three events, with the block timestamp of each:
\begin{center}\footnotesize
\texttt{OrderFilled(bytes32 orderHash, address maker, address taker, uint8 side, uint256 tokenId,}\\
\texttt{\quad uint256 makerAmountFilled, uint256 takerAmountFilled, uint256 fee, bytes32 builder, bytes32 metadata)}\\
\texttt{OrdersMatched(bytes32 takerOrderHash, address takerOrderMaker, uint8 side, uint256 tokenId,}\\
\texttt{\quad uint256 makerAmountFilled, uint256 takerAmountFilled)}\\
\texttt{FeeCharged(address payer, uint256 amount)}
\end{center}
On the first-generation exchange contracts the fill event is
\begin{center}\footnotesize
\texttt{OrderFilled(bytes32, address, address, uint256 makerAssetId, uint256 takerAssetId,}\\
\texttt{\quad uint256 makerAmountFilled, uint256 takerAmountFilled, uint256 fee)}
\end{center}
without a side; the side is a buy when the maker's asset is the collateral (asset id zero) and a
sell otherwise, and the token is the other asset.

\paragraph{The taker leg.} A settlement emits one \texttt{OrderFilled} per order it fills. The leg
whose \texttt{taker} field is the exchange contract itself is the taker order; every transaction
has exactly one. On the second-generation exchange the \texttt{OrdersMatched} event repeats that
leg's side, token and amounts, and agrees with it on every field on every transaction we checked.
For a buy the taker paid \texttt{makerAmountFilled} of collateral for \texttt{takerAmountFilled}
shares, so the price is their ratio and the size the share amount; for a sell the two roles swap.
Both are in units of $10^{-6}$.

\paragraph{Settlement types.} With the taker leg on token $a$ and side $s$, every maker leg is
either on $a$ with side $-s$ or on the complement of $a$ with side $s$. A transaction whose maker
legs are all of the first kind is a complementary match; all of the second kind with $s$ a buy is
a mint and with $s$ a sell a merge; a transaction with both kinds is mixed. The complement of a
token is the other token of the same condition, from the venue's market metadata. A transaction
with a maker leg on a token that the metadata map does not contain is reported as unknown; on
every day every such token is one the archive never carried, so the class measures the archive's
market coverage, not a third token.

\paragraph{Trace for the observed signing error.}
The print illustrated in Section~\ref{sec:method-signed} is the second physical row of the
deposited \texttt{prints/day=2026-07-17.parquet} cache. Its transaction is
\href{https://polygonscan.com/tx/0x957409201589d6ca83e168adcc2e9f56364350b4cc5d50504bf46df51bbedbb5}{\nolinkurl{0x957409201589d6ca83e168adcc2e9f56364350b4cc5d50504bf46df51bbedbb5}}.
Block 90{,}398{,}405 records the settlement at 15:36:45 UTC, after the print's receipt.
The taker leg buys 115.07 shares at 0.22; two maker legs buy 15.07 and 100 shares of the
complement token. The token map places both tokens in the same condition with opposite
outcomes, so this is a mint: all three order legs are buys, but only the taker leg carries
the print's token. A quote received in the print's millisecond is excluded under the timing
rule; the example uses the strictly preceding quote. This case was chosen to illustrate the
signing mechanism and is not an estimate of its prevalence. Fees are excluded from the
spread calculation.

\paragraph{Block ranges.} Each day's first and last block are found by binary search on block
timestamps, and the range is extended by 300 blocks (seven and a half to ten minutes) past
midnight so that the settlements of the day's last prints are read; those tail settlements are
excluded from the count of settlements whose print the archive does not contain. Polygon's block
time was cut in two steps over the sample, from 2.0 to 1.75~seconds in early May and to
1.5~seconds by early June, so a block count per day is not a constant. For an hour-stratified
sample of the unmatched prints, the transaction receipt from \texttt{eth\_getTransactionReceipt}
puts each print in one of five classes: no receipt (the hash was never mined), a receipt with
status zero (reverted), a block outside the range, a recipient other than the exchange contracts,
or a success on an exchange inside the range, which would be a gap in the scrape. The coverage
table reports the first two as ``no tx'' and ``reverted'' and pools the other three as ``other''.

\section{Archive health}
\label{app:health}

\begin{figure}[htbp]\centering
\includegraphics[width=\linewidth]{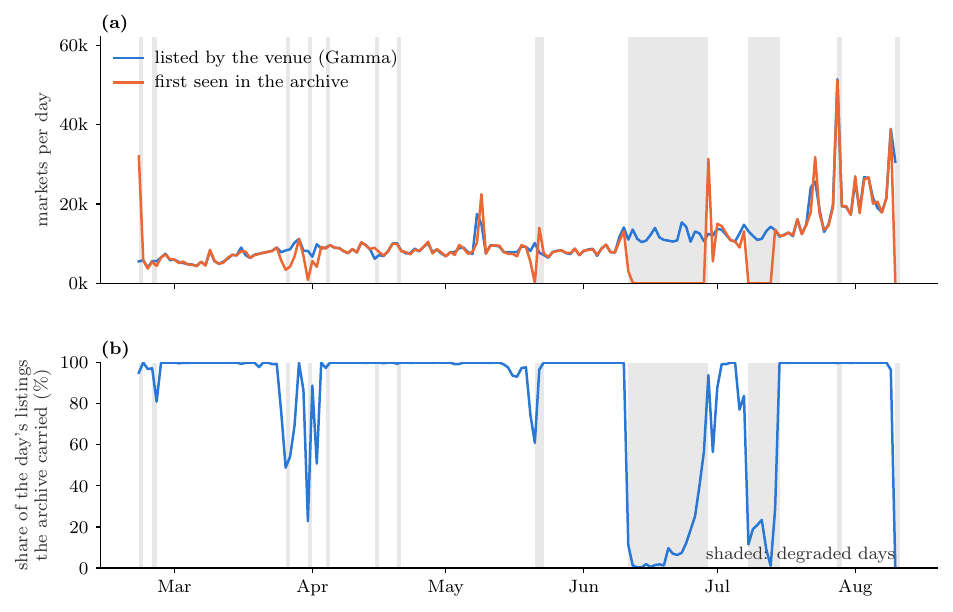}
\caption{Archive coverage of the venue's listings, by day. Panel~(a): markets listed by the venue,
from a sweep of its metadata API by listing day, and markets first seen in the archive; catch-up
days after an outage exceed the axis. Panel~(b): the share of each day's listings the archive
carried on that day or later. Shaded: the degraded days of Table~\ref{tab:health}.}
\label{fig:health}
\end{figure}

\paragraph{The benchmark.} The venue's listings are taken from a sweep of its metadata API by
listing day, 1{,}933{,}693 markets listed between 14 February and 12 August, and not from the
markets the archive carried: a count built from the archive collapses in exactly the windows the
collector missed, because a market listed and resolved inside an outage was never carried and so
never enumerated. On the 171 archive days the venue listed 1{,}858{,}106 markets and the archive
carried 1{,}511{,}091 of them (\SI{81.3}{\percent}) on the listing day or later. On the clean days,
those that pass the rule below and so carry at least half of their listings, the archive carried
\SI{93.6}{\percent} of the day's listings on the tenth-percentile day and \SI{99.98}{\percent} on
the median day, and its count of markets first seen tracks the venue's listing count day by day
(Figure~\ref{fig:health}a).

\paragraph{The rule.} A day is degraded when (i) one of its hourly files is missing or holds under a
fifth of the median hour's 75.6~million rows, (ii) the number of assets first seen in the archive
is below a tenth of the median over the previous seven non-degraded days while the venue's listings
are at least half their median over those days, or (iii) the archive carried fewer than half the
markets the venue listed that day. The median in (ii) runs over non-degraded days so that the rule
keeps flagging an outage longer than a week. The provider's receipt files, which record each
hourly file, exist only for the archive's second format, so the near-empty test of (i) applies only
to the second-format files and the missing-file test to both formats. The near-empty 11:00 and
12:00 files of 28 April are exempt from (i): the first falls in the
downtime the venue announced (Section~\ref{sec:migration}) and the second just after it, as the books reopened without the wiped orders
(Section~\ref{sec:cutover} shows the empty books at 12:00). The thresholds were fixed before the
rule was run, at one fifth, one tenth, one half and seven days, and three of them are
varied: moving the first-seen ratio to one third changes nothing, moving it to two thirds adds
three days (27 March, 2 April, 30 June), moving the carried share to one third removes one day
(26 March) and to two thirds adds the same three, and a five- or ten-day window changes nothing.
The one-fifth threshold of (i) and the listings condition of (ii) are not varied. The rule has a
blind spot it cannot close: it benchmarks listings by listing day, so it catches a collector that
misses new markets but not one that drops older markets still trading. The two on-chain days of
Section~\ref{sec:identity} on which the archive lacked whole markets or delivered late prints,
5 June and 1 July, pass the rule, and the on-chain coverage is the check the rule does not make.
Thirty-six of the 171 days are degraded.

\begin{table}[htbp]\centering\footnotesize
\caption{Degraded days. Rule: the test that flags the day, (i) a missing or near-empty hourly
file, (ii) almost no assets first seen while the venue keeps listing, (iii) under half of the day's
listings carried.}
\label{tab:health}
\begin{tabular}{llp{8.6cm}}
\toprule
Days & Rule & Signature \\
\midrule
11--28 June & (ii), (iii) & Subscription list frozen for eighteen days: no market first seen on
fifteen of them; assets observed decay from 82{,}722 to 31{,}006 as old markets resolve; the venue
lists 10{,}399 to 15{,}365 markets a day; of the markets listed 13--19 June the archive ever carried
\SI{0.3}{\percent} to \SI{1.9}{\percent}. Refreshed on 29 June, when 31{,}376 markets appear.
Three hourly files are missing on 11 June, which (i) also flags. \\
8--14 July & (ii), (iii) & The same signature over seven days: 0 to 66 markets first seen on six of
them, then 13{,}537 on 14 July, a day on which the archive still carried \SI{28.8}{\percent} of the
venue's listings. \\
21 Feb, 10 Aug & (i) & First and last archive days: 8 and 1 hours. \\
24 Feb, 4 Apr & (i) & 22 and 23 hours. \\
15 Apr & (i) & The seam: 9 hours of the archive's first format before the second begins. \\
20 Apr, 22 May, 28 Jul & (i) & One or two near-empty hourly files. \\
26 Mar, 31 Mar & (iii) & 3{,}426 and 876 markets first seen against 8{,}304 and 8{,}194 listed;
\SI{48.8}{\percent} and \SI{22.8}{\percent} carried. \\
21 May & (ii) & 102 markets first seen against 10{,}185 listed; the catch-up (14{,}017 markets)
came on 22 May. \\
\bottomrule
\end{tabular}
\end{table}

\paragraph{The venue's listing rate.} The rate at which the venue lists markets is itself not
constant: a median of 8{,}759 markets a day from February to July and 23{,}547 in August, with a
peak of 51{,}477 on 28 July. A count of assets in the archive therefore grows over the sample for
a reason that has nothing to do with the collector, and Section~\ref{sec:composition} reads the
growth against this benchmark. The archive's first format carries no trade prints, so trade
counts exist only for its second-format days.

\section{Venue changes and supporting liquidity analyses}
\label{app:venue-analysis}

These analyses provide institutional and data-quality context for the direction benchmark.
They describe venue events and quoted liquidity; they do not identify a causal effect of
trade-sign errors, migration or the tick.

\subsection{The price rule and the dated changes}
\label{sec:venue-changes}

Markets move from a 0.01 tick to a 0.001 tick one by one, each on its own date, both of a
market's assets within a minute of each other. The venue's
documentation of the \texttt{tick\_size\_change} message stated, as of June 2025, that the message
is emitted when the book's price reaches the limits, above 0.96 or below 0.04; the page as of
September 2026 documents the message and its two values without that sentence
\citep{polymarket-docs-market-channel}. We call that trigger the \emph{price rule}, and
Section~\ref{sec:tick} checks it against the events.

The analysis uses a list of dated changes, here in date order. On 6 March taker fees and maker
rebates were extended to all crypto markets, a change announced in the venue's changelog of 1 March
that applied only to markets created after 6 March. On 30 March the venue changed its
fee schedule (its ``Fee Structure V2''), without an order wipe. The archive's provider began its
second format at 19:00 UTC on 13 April \citep{pmxt-archive-v2}, and the archive this paper uses
switches to it on 15 April; neither is a
venue event. The venue migrated to V2 on 28 April, and on 28 May its taker rebate programme went
live. On 2 July the World Cup moneyline, spreads, totals and to-advance markets moved to a 0.0025 tick, a
change that the markets' own prices did not trigger; we call it the World Cup change. On 10 July
the sports taker fee rate, a coefficient in the venue's fee formula, rose from 0.03 to 0.05, and
the sports maker rebate fell from 25 to \SI{15}{\percent} of the taker fees collected. On 17 July
the venue announced a change that took effect on 24 July: the order endpoints stopped returning
transaction hashes for immediately matched (fill-and-kill and fill-or-kill) orders, which are
retrieved by polling, and matching latency was cut. The last change
on the list, on 7 August, altered how crypto up/down markets resolve. The migration is the event
that Section~\ref{sec:method-cutover} describes, and its placebo windows are kept away from every
date on the list; the 30 March fee change and the archive's format change also serve there as
comparisons. The World Cup change is this paper's separate tick comparison.

\subsection{Related evidence on technology and tick-size changes}
\paragraph{Technology change and market quality.}
The 28 April migration is a venue-side technology change with a dated cutover, the setting of
\citet{hendershott-moulton-2011-hybrid} and \citet{riordan-storkenmaier-2012-latency}. Those
studies had a comparison group of unaffected securities or venues; a platform that migrates every
market at once has none, and we rank the change among placebo dates instead.

\paragraph{Tick size and staggered treatment.}
Identifying an effect of the tick requires an assignment set outside the market and an adoption
unrelated to the outcome's trend. The tick-size experiments that identify cleanly meet the first
requirement: the randomised US Tick Size Pilot \citep{rindi-werner-2019-pilot}, the dated
introduction of the MiFID II tick-size regime \citep{eu-2017-588-tick} and the venue competition that let the same stock
trade on different grids \citep{foley-2023-tickwars}. Tick size shapes quoted spreads, depth and
the incentive to supply liquidity \citep{harris1994minimum}; in the equity reductions to sixteenths
and to decimals, quoted spreads narrowed \citep{goldstein2000eighths,bessembinder2003trade} and
depth at the best quote fell \citep{goldstein2000eighths}. Many venues set the tick by price band,
so a tick that depends on price is not exotic; on a binary market the top band has an implied
probability near one, and on this venue the tick changes turn out to be that rule. Because markets
change tick one by one rather than all at once, the changes invite a staggered design. The recent
econometrics of staggered adoption shows that two-way fixed-effects and pooled comparisons are
biased when effects are heterogeneous and already-treated units serve as controls
\citep{goodmanbacon2021difference,dechaisemartin2020two,sun2021estimating,borusyak-2024-revisiting},
and that not-yet-treated or never-treated units are the appropriate comparison
\citep{callaway2021difference,baker-2022-staggered}. That comparison still needs the second
requirement, and an adoption that follows the price does not meet it, so the staggered event study we report cannot identify a tick effect. The World Cup change does not depend on the
market's price; it is small, and we treat it separately.

\FloatBarrier
\subsection{Composition of the quoted-spread panel}
\label{sec:composition}

\begin{figure}[htbp]\centering
\includegraphics[width=0.9\linewidth]{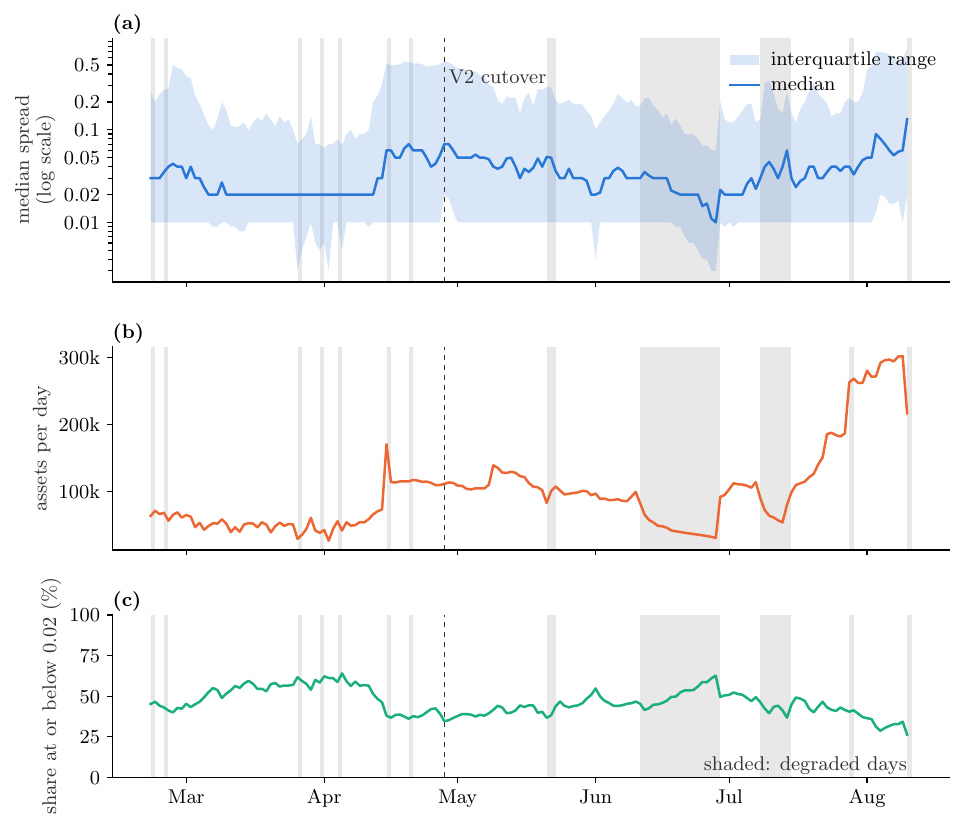}
\caption{The panel over both archive formats, 21 February to 10 August. Panel~(a): the daily
cross-section of the median spread over live asset-days, median and interquartile range, on
a log scale. Panel~(b): assets observed in the archive each day. Panel~(c): the share of live
asset-days with a median spread at or below 0.02. Shaded: degraded days; the dashed line is the
28 April cutover. The venue's own listing rate is in Figure~\ref{fig:health}.}
\label{fig:series}
\end{figure}

\begin{figure}[htbp]\centering
\includegraphics[width=0.62\linewidth]{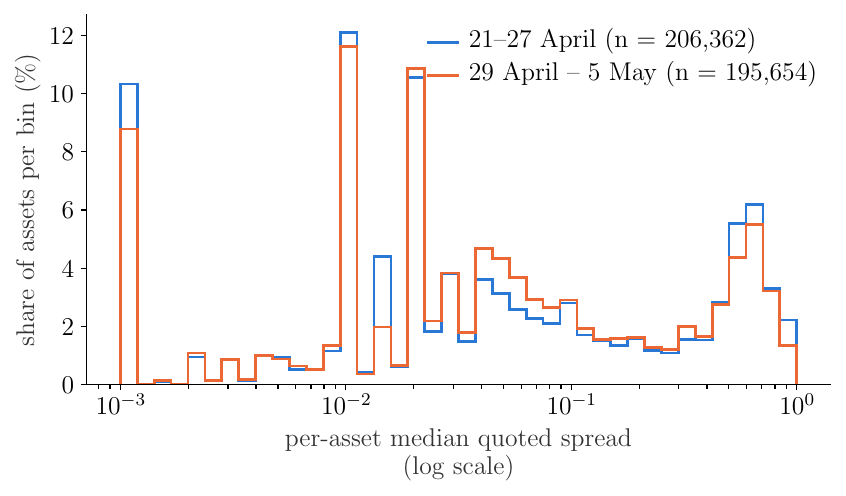}
\caption{Distribution of per-asset median quoted spreads over live asset-days in the week before
and the week after the cutover, both in the archive's second format. The cross-section is bimodal,
with mass at a tight mode near one tick and a wide mode above 0.5, so a before/after difference is
a mixture change as much as a level change; this is why Section~\ref{sec:cutover} conditions on a
cohort.}
\label{fig:dist}
\end{figure}

Figure~\ref{fig:series} shows why no whole-panel series in this paper is read as a quality series.
The number of assets the archive observes a day rises from 71{,}802 on 22 February, the first clean
day, to 302{,}256 on 9 August, the last. Over the same months the venue's listing rate rose from a
median of 8{,}759 markets a day between February and July to 23{,}547 in August, so most of the
growth is the venue's. The rest is consistent with a wider subscription in the archive's second
format: across the seam of 15 April, 65{,}646 assets were live on 14 April in the first format and
103{,}726 on 16 April in the second, 40{,}408 of them on both days. The spread cross-section is
bimodal, a tight mode at one or two ticks and a wide mode above 0.5 (Figure~\ref{fig:dist}), and on
clean days its daily median over live asset-days moves between 0.020 and 0.090 while the share of
live asset-days at or below 0.02 moves between \SI{29}{\percent} and \SI{64}{\percent}. Across the
cutover, 97{,}889 assets were live on 27 April and 92{,}892 on 28 April, 78{,}763 on both; the
cohort of Section~\ref{sec:cutover}, live on each of the seven days before, is about half the 27
April count. Every comparison in this paper is therefore within assets or against a benchmark that
shares the composition, and the venue's listing rate is the benchmark for growth.

\FloatBarrier

\subsection{Additional archive diagnostics}
Two windows show the signature of a collector whose subscription list stopped refreshing: from 11
to 28 June and from 8 to 14 July (the June and July outages) the number of assets observed a day
decays while new markets stop appearing, although the venue listed 10{,}000 to 15{,}000 markets a
day throughout. Three hours are missing on 11 June, and six hourly files hold under a fifth of a
normal hour's rows, among them the 11:00 file of the cutover day, 28 April, which holds 1.2~million
rows against a median of 75.6~million; that file and the next are the venue's migration outage and
do not make the day degraded. The tick classification, the adoption hazard and the two
health-screened and hazard-reweighted event-study specifications drop the degraded days, and the
cutover and World Cup windows contain none; a placebo window may contain degraded days outside its
ranked days, while the baseline event-study samples and the 30 March comparison keep them. The
two on-chain days with the lowest feed coverage in Table~\ref{tab:fillmatch}, 5 June and 1 July,
fall six days before and three days after the June outage; Section~\ref{sec:identity} shows what
the archive missed on them.

\subsection{Cutover comparison: methods}
\label{sec:method-cutover}

The migration of 28 April moved every market at once, so there is no contemporaneous control
group. We therefore describe it within assets and place it among other dates. The cohort is the
set of assets that were live on each of the seven days before the event, a live day being an
asset-day whose median spread is above zero and below 0.9, the cut at which a quote counts as
empty (Section~\ref{sec:data}). For each cohort asset and each event day $k\in[-7,7]$ the outcome
is the change from the day before the event, $\Delta_k = y_{k} - y_{-1}$, and the paper reports
the cohort median and mean of $\Delta_k$ with an asset bootstrap of the median (1{,}000 draws),
together with the net share of assets whose outcome rose, the share up less the share down. The
outcomes are the log of the median quoted spread, the log of the update count, the log of the mean
displayed size at the best bid from \texttt{price\_change} rows at the best level and the
empty-quote share, the share of the day's updates that show an empty quote. An asset drops out of
a day's pair when it has no update that day or its outcome is undefined, as the log of a zero
median spread is. A day on which every update shows an empty quote stays in the spread's pair, so
the spread's paired change carries the emptied books the feed still updated along with the assets
that kept a quoted book. The empty-quote share and the hour-by-hour path follow the emptied books
directly, and an adverse-tail sensitivity at $k=7$ assigns each leaver, a cohort asset whose outcome is
missing on that day, the adverse tail of the cohort's pre-period distribution: the 90th percentile
of an outcome whose rise is adverse (spread, empty-quote share) and the 10th of one whose fall is
(updates, displayed size). This percentile imputation is a sensitivity scenario, not a
mathematical worst-case bound on the missing outcomes.

The inference is a ranking among placebo dates. The same cohort rule and the same paired changes
are computed with every other date of the panel as the event. A date qualifies when its window is
inside the panel and holds no run of three or more degraded days, when none of the days whose
changes are ranked ($k=-1$ to $3$) is degraded and when the date is more than seven days from every
date on the analysis's list of dated changes in Section~\ref{sec:venue-changes} (6 and 30 March;
13 and 15 April; 28 April; 28 May; 2, 10 and 24 July; 7 August), so that no placebo window contains
one of them; 21 dates qualify. The rank of 28 April among the 22 is the paper's
statement about the cutover. The directions were fixed before the data were seen (spread up,
updates down and depth down, with a joint rank on spread and updates); the two-sided ranks by
absolute size are post hoc and exploratory, and both are reported. The dates overlap in their
windows, adjacent dates sharing fourteen of fifteen days, so
three subsets are reported as checks: 13 dates whose whole window holds no degraded day, 12 dates
from the archive's second format and 2 dates spaced fifteen days apart from each other and from
the event, whose windows share no day with each other or with the event's; that subset's smallest
attainable rank p-value is one in three. Because the cohort median asset's spread is pinned to a
tick, the median change is exactly zero on ordinary days and its rank is uninformative on its own;
the mean and the net share are the statistics that can rank the event against a distribution.

Further comparisons accompany the ranks, and none of them attributes the change to one of the
bundled design changes. The hourly path follows the cohort through the cutover week hour by hour,
so that a rise confined to the hours after the order wipe can be told from anticipation. The
splits divide the cohort by pre-period activity quartile, by age at the event and by venue
category, and the category difference-in-differences of days $+5$ to $+7$ against $-7$ to $-1$
is reported next to each category's taker fee rate. The archive's own format change on 15 April,
which is not a venue event, is put through a paired change from 14 to 16 April on the assets live
on both days, as a placebo for the measurement seam. The 30 March fee change, a dated venue change
without an order wipe, is put through the cutover's paired change as a comparison. Last, the
displayed size at the best bid is cross-checked against the book snapshots the feed sends when it
subscribes, because a migration can change what a \texttt{price\_change} size means as well as
what the book holds.

\subsection{Tick-change comparisons: methods}
\label{sec:method-tick}

Section~\ref{sec:venue-changes} gives the price rule as the venue's documentation stated it. We
check the rule against the events instead of assuming it, and report the exceptions. Each of the
0.01-to-0.001 events in the panel is joined to the last \texttt{price\_change} of the same asset
within the preceding 60 minutes on the same UTC day and to the venue's metadata for its market, and
takes the first of four classes it satisfies. The classes overlap, so their shares depend on the
order, and the overlap is reported. The classes are a price-rule class, when that quote is
live (its recorded spread below 0.9, a missing side counted at 0 or 1) and its midpoint is above
0.96 or below 0.04; an \emph{activation} class, when the event falls within 24 hours of the asset's
first appearance in the archive, or before its first live quote, or before its first trade; an
\emph{operator batch} class, when at least 50 tick changes fall within 60 seconds in one group of related
markets (an `event' in the venue's metadata); and a residual. For each class we report where the
quote sat, whether the market's other token changed tick within a minute and how far the event lies
from the market's closing time and, for sports markets, from the game's start. An asset's first such
change is its adoption. A discrete-time hazard of adoption on the asset-day's state the day before
(update rate, trading, placeholder book, price extremeness, spread at the 0.01 floor, age,
neg-risk), fitted on all adoptions and a five-percent sample of non-adopting asset-days, supplies
the weights for one further specification of the event study. The events of 2 July, which move World
Cup markets to a 0.0025 tick, are a separate class and a separate design.

\paragraph{Event study.} The group-time event study uses an auxiliary contract-day panel
over the second-format days from 15 April to 10 August. Its outcome is the day's mean quoted
spread, distinct from the harmonised panel of Section~\ref{sec:data}. Adoption is the asset's
first 0.01-to-0.001 tick change; extraction timestamps stored in Europe/Zurich wall time are
converted to UTC on read. The event study counts adopting assets, whereas the event
classification counts token-level changes on clean days. For
cohort $g$ (adoption day) and day $t$, $\mathrm{ATT}(g,t)$ is the mean change in the log of the
day's mean quoted spread from day $g-1$ to day $t$ among cohort $g$, less the same change among
assets that adopt after $t$ or not at all, in the manner of \citet{callaway2021difference}; cohort
effects are aggregated by event time $e=t-g\in[-7,7]$, weighted by cohort size, in three samples (a
fixed one-in-five sample of all assets; adopters and controls that traded on day $g-1$; adopters
already trading whose update rate changed by less than a quarter across adoption, against the
controls of the second sample). Pre-period effects ($e\le-2$) test parallel trends with a bootstrap
over adoption cohorts (300 draws). Two sensitivity specifications address archive health and
selection. One removes the archive's degraded
days from the panel, and with them the cohorts whose adoption day or base day is degraded. The
other, on that same sample, weights the control units by the odds of their fitted adoption hazard,
$p/(1-p)$, averaged over each asset's days in the panel of Section~\ref{sec:data} and capped at 20,
so that the controls likelier to adopt count for more. The estimates are reported as descriptions of
what happens to a spread around the event; Section~\ref{sec:tick} explains why none of them is an
effect of the tick.

\paragraph{The World Cup change.} On 2 July at 00:31 UTC the venue moved the World Cup markets to a
0.0025 tick: one arm from 0.01 (a finer grid) and one from 0.001 (a coarser one). It is the only
tick change in the sample that the market's own price did not trigger, and the two arms predict
opposite signs for what the grid sets mechanically, the spread where the grid binds and the size a
best level holds. A state common to both arms, such as a game reaching its decision, moves both the
same way, which is why the controls are matched on the game's timing. Controls are the venue's other
sports markets of the same types by the venue's own field (moneyline, spreads and totals; the treated
markets also include to-advance markets). A control must have been quoted on 1 July, its game must
start at least half a day after the change and before the end of the window (2 July 12:31 to 3 July
24:00 UTC), its tick before the change must match the arm's (the last tick event before it, else the
0.01 tick every market is listed with) and it must pass a pre-period spread screen (a median
below 0.5 of the median spreads of its pre-period bins, both tokens pooled, from at least two
token-bins). This excludes wide books but does not prove that both sides are present: a
one-sided book near resolution can also have a narrow recorded spread. In the finer arm a control must also be on the listing grid: 23 candidate controls had
a pre-period half-day median spread below 0.0095, which puts them on a finer grid than the listing
tick assigned to them, and are excluded. The outcomes are the log of the half-day median spread, the
spread in old ticks, the share of quotes within one old tick and the displayed size at the best bid,
in twelve-hour bins from 30 June 12:00 to 3 July 24:00. The first post bin starts at
00:00 on 2 July and includes 31 minutes before the change. The estimate is a two-way fixed-effects
difference-in-differences with standard errors clustered by market, beside the paired median change
and an exploratory permutation of market labels (2{,}000 draws, with a plus-one correction).
The minimum detectable effect at the arm's size, from the spread of the control markets' own
pre-to-post changes, is a nominal calculation under independent markets, not a power guarantee.
Contracts on the same game share shocks. We therefore also cluster the two-way estimate by the
venue's event identifier and bootstrap whole events within treated and control groups, retaining
every contract of a resampled event. Both arms are descriptive: World Cup status is not random,
there are few treated events, and parallel trends cannot be established in this short window.
The market-label permutation does not reproduce the venue's assignment mechanism and has no
causal randomisation interpretation; its p-value is shown only as an exploratory reference
where the permutation distribution is not degenerate.
 
\subsection{Inference for supporting analyses}
The cutover's inference is the rank of 28 April
among placebo dates (Section~\ref{sec:method-cutover}); the dates overlap in their windows, so the
count of dates overstates the number of independent draws, and ranks are also reported on dates
spaced a window apart; its within-cohort intervals are an asset bootstrap of the median (1{,}000
draws). Adoptions on the same day share a control change, so the event-study intervals and the
joint pre-trend test come from a bootstrap over adoption cohorts (300 draws). The World Cup arms
carry a market-clustered standard error on the two-way fixed-effects estimate, a market bootstrap
of the paired estimate (1{,}000 draws), with event-clustered standard errors and a bootstrap
of whole events as sensitivity checks. Market-label permutations are exploratory references
without a causal randomisation interpretation; both World Cup arms are descriptive.

\subsection{Cutover comparison: results}
\label{sec:cutover}

\begin{table}[htbp]
\centering\scriptsize\setlength{\tabcolsep}{2.5pt}
\begin{tabular}{lrrrrrrrrrrrr}
\toprule
Outcome & $\bar\Delta_{-7}$ & net$_{-7}$ & $\bar\Delta_{-3}$ & $\tilde\Delta_0$ & $\bar\Delta_0$ & net$_0$ & $\tilde\Delta_1$ & $\tilde\Delta_3$ & $\tilde\Delta_7$ & half-life & survivors & tail scenario \\
\midrule
log spread & 0.541 & 0.452 & 0.319 & 0.016 & 0.309 & 0.314 & 0.037 & 0.000 & 0.000 & 2 & 34{,}496 & 0.106 \\
log updates & $-0.171$ & $-0.080$ & $-0.244$ & 0.023 & 0.070 & 0.026 & $-0.147$ & 0.123 & $-0.007$ & 7 & 34{,}496 & $-0.359$ \\
log depth, best bid & $-0.315$ & $-0.117$ & $-0.351$ & 0.519 & 0.625 & 0.406 & 0.294 & 0.374 & 0.270 & 5 & 28{,}036 & $-0.296$ \\
empty-quote share & 0.003 & 0.011 & 0.000 & 0.050 & 0.124 & 0.951 & 0.000 & 0.000 & 0.000 & 1 & 34{,}496 & 0.000 \\
\bottomrule
\end{tabular}
\caption{The 28 April cutover: paired within-asset changes from 27 April on the cohort of 51{,}560 assets live on each of the seven days before. $\Delta_k$ is the change on day $k$ from day $-1$: $\bar\Delta$ is the cohort mean, $\tilde\Delta$ the cohort median and net the share of assets whose outcome rose less the share whose outcome fell; the text gives an asset bootstrap of the median at $k=0$. The empty-quote share is the share of the day's updates that show an empty quote. The half-life is the first day after the event on which the median change, in absolute value, is at most half its value on the event day. Survivors are the cohort assets whose outcome is observed on both days of the pair at $k=7$, and `tail scenario' is the median change on that day under a sensitivity scenario that assigns each leaver the adverse tail of the cohort's pre-period distribution (the 90th percentile for spread and the empty-quote share, the 10th for updates and depth). Table~\ref{tab:placebo} ranks the changes among placebo dates.}
\label{tab:cutover}
\end{table}
 
\begin{figure}[htbp]\centering
\includegraphics[width=\linewidth]{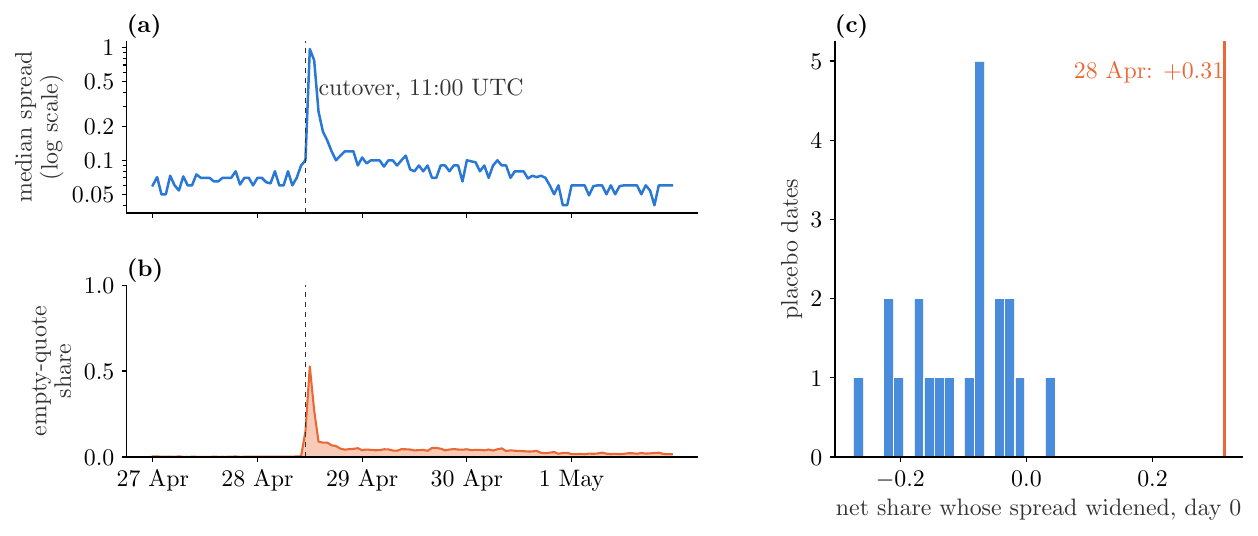}
\caption{The cutover on the cohort of assets live on each of the seven days before. Panel~(a):
the cohort's hourly median spread, with empty quotes counted at their recorded spread; panel~(b):
the share of quoted cohort assets whose last quote of the hour was empty; both from 27 April to 1
May, the dashed line at 11:00 UTC on 28 April. Panel~(c): the net share of cohort assets whose
spread widened on the event day (the share that widened less the share that narrowed), over 21
placebo dates, with 28 April marked.}
\label{fig:cutover}
\end{figure}

Table~\ref{tab:cutover} and Figure~\ref{fig:cutover} describe the migration on the 51{,}560 assets
that were live on each of the seven days before it. For the asset in the middle of the cohort, the
day of the cutover was a small event. The median change in log spread is 0.016 on 28 April and
0.037 on 29 April, with asset-bootstrap intervals of $[0.0155, 0.0171]$ and $[0.0333, 0.0465]$, and
it is back at zero by 30 April, a half-life of two days. The mean change is larger, 0.309 on the day,
and the net share of assets whose spread widened is 0.314: 0.51 of the cohort widened and 0.198
narrowed, five for every two. Both lie outside the range of every placebo date. The empty-quote
share rose by 0.050 for the median cohort asset on the day, and its cohort mean was still 0.048 on
29 April and 0.02 a week later.

The cohort was not standing still. Its mean log spread relative to 27 April runs from 0.541 seven
days before to a third of a log point three days before and 0.154 two days before, so spreads were
narrowing into the event: 0.625 of the cohort had a wider spread a week before than on 27 April
and 0.173 a narrower one. The day's mean change reverses that drift. A third of the cohort left the
panel within the week after; the rule that selects assets with a week of live quotes selects
markets heading toward their decision. By 5 May, 34{,}496 of the cohort still had quotes with a
positive median spread; with every leaver assigned the cohort's pre-period 90th percentile spread,
the median change on that day is 0.106 instead of zero. The same scenario is uninformative for the
empty-quote share, whose pre-period 90th percentile is zero; the leavers include the emptied
books, and their worst case is one.

\paragraph{Hour by hour.} Before 11:00 on 28 April the cohort's median spread was close to its
level at the same hours of 27 April but not identical: the hourly log ratio between the two days has
a median of 0.105 over the morning hours, against 0.357 at 11:00 and 2.63 at 12:00. At 12:00,
\SI{53}{\percent} of the cohort assets with a quote in that hour ended it on an empty quote, and the
median spread, which counts an empty quote at its recorded spread, was 0.97; at 13:00 the share
was \SI{27}{\percent}, at 15:00 \SI{8}{\percent} and from 19:00 it stayed between
\SI{3.6}{\percent} and \SI{5.2}{\percent} through 29 April. The median spread was 0.18 at 15:00
and 0.10 at 18:00, and on 29 April it ran between 0.065 and 0.11 against 0.05 to 0.08 on 27 April. This
is the signature of the order wipe: little before the hour, empty books at the hour and a residual
of books that were not reposted. The hourly medians are cross-sectional and include those empty
books, while the daily paired changes in Table~\ref{tab:cutover} fold an asset's empty updates
into its median for the day and drop the assets with no positive median that day; that is why the
two describe the same week with different magnitudes. Cross-sectional medians also pool
emptied books, whose spread is recorded at or near one, with the rest, and new listings with the
cohort; they do not measure the same within-asset change.

\paragraph{Displayed size.} The rise in displayed size at the best bid is the only result that the
feed's own change contaminates. The book snapshots the feed sends on subscription give a second
measure of the same quantity, on another sample: the cross-check takes every asset quoted on both
days, Table~\ref{tab:cutover} the cohort, and the two are not meant to agree. On the day of the
migration both measures rose, the median by 0.475 from the \texttt{price\_change} rows and by
0.372 from the snapshots. After the day they part: on 29 April the \texttt{price\_change} measure
was still 0.268 above its 27 April value and the snapshot measure 0.069; on 5 May, 0.32 against
0.15. What a \texttt{price\_change} size reports relative to the book changed around the
migration, and the depth outcome is therefore not comparable across it; the ratio of the two
measures had already moved on 25 and 26 April, when snapshot coverage was thin, so the date of the
change is not sharp either. The snapshot figures are the most that can be said. Displayed size at
the best bid rose on the day and was 0.051 to 0.171 log points higher on each day of the following
week, which is of the same size as the cohort's own pre-period drift in median depth ($-0.114$ and $-0.122$ at $k=-7$ and
$-3$); the drift could explain it.

\paragraph{Where 28 April ranks.} Table~\ref{tab:placebo} places the event among 21 other dates. On
the event day the mean change in log spread, 0.309, lies outside the placebo range of $-0.259$ to
$-0.020$; the net share widening, 0.314, outside $-0.274$ to 0.046; the empty-quote share's net
share up, 0.95, outside a range whose maximum is 0.1734; and the median change in displayed size at
the best bid, 0.519, outside a range whose maximum is 0.0456. Each is rank 1 of 22, two-sided. The
pre-specified joint rank on spread up and updates down is 1 of 22, but it is taken on the day-0
medians and inherits the spread median's degeneracy (no placebo date can match a nonzero median);
the same joint rank on the mean and on the net share is also 1 of 22, and the update count on its
own ranks 11 of 22 in the pre-specified direction (17 of 22 two-sided). In the pre-specified
direction the displayed size ranks 22 of 22: it rose where the design expected a fall, and the
two-sided rank is exploratory and post hoc. The median change in log spread is rank 1 too, but
only because it is exactly zero on every other date. On the day after the event the mean change, 0.214,
still lies above every placebo date, whose day-1 means are all negative, but seven of them move
further in absolute size, so its two-sided rank is 8 of 22; the net share widening is still in the
tail (rank 3 of 22). The update count is not distinguishable from an ordinary day on either day (ranks 12 and
21 of 22 for the mean); its median fall of 0.147 on 29 April is matched by the ordinary movement
between weekdays, which the placebo dates contain. On the 13 dates whose whole window holds no
degraded day, the 12 dates of the archive's second format and the 2 dates spaced a window apart
from each other and from the event, the mean and the net share on the event day rank first in
each; the spaced subset can attain no p-value below one in three, so it checks overlap and nothing
more.

\begin{table}[htbp]
\centering\scriptsize\setlength{\tabcolsep}{2pt}
\begin{tabular}{lrrrrrrrrrrrrr}
\toprule
Outcome & \multicolumn{6}{c}{day 0} & \multicolumn{6}{c}{day 1} & strict \\
\cmidrule(lr){2-7}\cmidrule(lr){8-13}
 & median & rank & mean & rank & net up & rank & median & rank & mean & rank & net up & rank & rank \\
\midrule
log spread & 0.016 & 1/22 & 0.309 & 1/22 & 0.31 & 1/22 & 0.037 & 1/22 & 0.214 & 8/22 & 0.29 & 3/22 & 1/14 \\
log updates & 0.023 & 17/22 & 0.070 & 12/22 & 0.03 & 18/22 & $-0.147$ & 9/22 & $-0.024$ & 21/22 & $-0.12$ & 12/22 & 8/14 \\
log depth, best bid & 0.519 & 1/22 & 0.625 & 1/22 & 0.41 & 1/22 & 0.294 & 1/22 & 0.448 & 2/22 & 0.29 & 1/22 & 14/14 \\
empty-quote share & 0.050 & 1/22 & 0.124 & 1/22 & 0.95 & 1/22 & 0.000 & 22/22 & 0.048 & 2/22 & 0.13 & 1/22 & 1/14 \\
\bottomrule
\end{tabular}
\caption{Where 28 April ranks among the 22 dates formed by the event and its placebo dates. For each outcome and statistic of the paired change from the day before (cohort median; cohort mean; net share of cohort assets whose outcome rose), the event's value and its two-sided rank by absolute size, rank 1 the most extreme. The strict column ranks the day-0 median among the 14 dates whose whole window contains no degraded day, in the pre-specified direction (spread up, updates and depth down) and, for the empty-quote share, upward; the text gives the strict ranks of the mean and the net share.}
\label{tab:placebo}
\end{table}

\begin{table}[htbp]
\centering\footnotesize\setlength{\tabcolsep}{4pt}
\begin{tabular}{lrrr}
\toprule
\multicolumn{4}{l}{\emph{A. Recovery by pre-period activity quartile (log spread)}} \\
quartile & assets & $\Delta_0$ & half-life \\
\midrule
q1 & 12{,}891 & 0.015 & 1 \\
q2 & 12{,}890 & 0.013 & 3 \\
q3 & 12{,}889 & 0.012 & 2 \\
q4 & 12{,}890 & 0.096 & 3 \\
\bottomrule
\end{tabular}
\vspace{4pt}

\begin{tabular}{lrrrrr}
\toprule
\multicolumn{6}{l}{\emph{B. Category difference-in-differences, days $+5$ to $+7$ against $-7$ to $-1$ (log spread)}} \\
category & assets & taker fee (\%) & treated & rest & difference \\
\midrule
sports & 21{,}752 & 3.0 & $-0.169$ & 0.000 & $-0.169$ \\
politics & 8{,}254 & 4.0 & 0.000 & $-0.087$ & 0.087 \\
culture & 2{,}112 & 5.0 & 0.010 & $-0.032$ & 0.042 \\
crypto & 1{,}644 & 7.0 & 0.067 & $-0.028$ & 0.095 \\
unknown & 1{,}190 & --- & 0.000 & $-0.019$ & 0.019 \\
finance prices & 1{,}030 & 4.0 & 0.000 & $-0.019$ & 0.019 \\
economics & 1{,}016 & 5.0 & 0.000 & $-0.018$ & 0.018 \\
tech & 844 & 4.0 & $-0.087$ & $-0.017$ & $-0.070$ \\
weather & 160 & 5.0 & $-0.490$ & $-0.017$ & $-0.473$ \\
general & 82 & 5.0 & 0.405 & $-0.017$ & 0.423 \\
mentions & 56 & 4.0 & $-0.019$ & $-0.017$ & $-0.001$ \\
\bottomrule
\end{tabular}
\vspace{4pt}

\begin{tabular}{lrrr}
\toprule
\multicolumn{4}{l}{\emph{C. The archive seam: paired change across the format change, no venue event}} \\
days & assets & median $\Delta$ log spread & within 10\% (\%) \\
\midrule
14 Apr $\to$ 16 Apr & 40{,}408 & $-0.018$ & 40.2 \\
\bottomrule
\end{tabular}
\caption{What the cutover is consistent with, on the cohort of 51{,}560 assets. Panel A splits the cohort by pre-period update-rate quartile (q4 the most active). Panel B compares each venue category's median change in log spread from the week before to days five to seven after with the rest of the cohort; the fee is the category's taker fee rate, the coefficient on $p(1-p)$, in percent. Panel C applies a paired change from 14 to 16 April across the archive's own format change on 15 April, on the assets live on both days, without a venue event.}
\label{tab:mechanisms}
\end{table}
 
\paragraph{What the change is consistent with.} The most active quarter of the cohort took the
largest hit on the day (Table~\ref{tab:mechanisms}), a median change of 0.096 in log spread against
0.015 in the least active quarter, and every quarter recovered with a half-life of one to three
days; the most active books had the most to lose from a wipe and took as long as any to be
repopulated. Split by age, which the table does not show, the 35{,}020 assets first seen in the
archive more than two weeks before the event recover with a three-day half-life, and the 16{,}540
newer ones have no half-life, because their median change on the day is zero; nearly half of them
leave the panel inside the week. Across the venue's categories, sports is the only large category
whose spread a week later was narrower than the rest's, by $-0.169$ on 21{,}752 assets. The taker fee
rate runs from 0.03 on sports to 0.07 on crypto; the venue's fee-free markets carry no fee schedule
in its metadata and sit in the unknown row, whose difference lies inside the range of the
fee-paying categories, so the fee model cannot be separated from the wipe in these data. The
archive's own format change on 15 April, put through a paired change on the 40{,}408 assets live
on both sides of it, moves the median asset's log spread by $-0.018$. The daily median effects of
the migration are of that order; what distinguishes the cutover from a measurement seam is the
hour-by-hour path, the emptied books and the tail of the cross-section, the large widenings that lift the mean
change to 0.309 against a median of 0.016. The 30 March fee change, on
a cohort of 19{,}342 assets, left the median spread unchanged on every day of its window and raised
the median update count by 0.129 on its third day and 0.310 on its seventh; 31 March is a degraded
day, so we draw nothing from it beyond the absence of a wipe signature.

\subsection{Tick-change comparisons: results}
\label{sec:tick}

\begin{table}[htbp]
\centering\scriptsize\setlength{\tabcolsep}{3pt}
\begin{tabular}{lrrrrrr}
\toprule
\multicolumn{7}{l}{\emph{A. Classes of the 1{,}551{,}795 tick events from 0.01 to 0.001, assigned in order}} \\
class & events & share (\%) & mid, high & mid, low & near threshold (\%) & paired (\%) \\
\midrule
price rule & 1{,}439{,}570 & 92.8 & 0.995 & 0.005 & 13.4 & 99.97 \\
activation & 71{,}795 & 4.6 & 0.900 & 0.250 & 0.0 & 99.97 \\
operator batch & 1 & 0.0 & --- & 0.100 & 0.0 & 100.00 \\
residual & 40{,}429 & 2.6 & 0.935 & 0.060 & 0.0 & 99.96 \\
\bottomrule
\end{tabular}
\vspace{4pt}

\begin{tabular}{lr}
\toprule
\multicolumn{2}{l}{\emph{B. The activation class against the asset's first live quote}} \\
quantity & value \\
\midrule
events in the activation class & 71{,}795 \\
share whose first live quote came before the event & 93.6 \\
share whose first live quote came within an hour after & 6.4 \\
median minutes from event to first live quote & $-776$ \\
\bottomrule
\end{tabular}
\vspace{4pt}

\begin{tabular}{lr}
\toprule
\multicolumn{2}{l}{\emph{C. Discrete-time hazard of adoption on the asset-day's state the day before}} \\
quantity & value \\
\midrule
area under the ROC curve & 0.747 \\
predicted daily hazard, placeholder book the day before & 0.173 \\
predicted daily hazard, first active day & 0.347 \\
\bottomrule
\end{tabular}
\vspace{4pt}

\begin{tabular}{lrrrrrrr}
\toprule
\multicolumn{8}{l}{\emph{D. The events against the market's closing time and, on sports markets, the game's start}} \\
class & events & one-sided (\%) & h to close & closed $<$24 h (\%) & with kick-off & h after kick-off & after kick-off (\%) \\
\midrule
price rule & 1{,}439{,}570 & 76.7 & 2.0 & 86.5 & 1{,}098{,}926 & 1.6 & 80.7 \\
activation & 71{,}795 & 0.0 & 3.4 & 75.1 & 59{,}304 & 0.7 & 65.1 \\
operator batch & 1 & 0.0 & 0.2 & 100.0 & 1 & 2.1 & 100.0 \\
residual & 40{,}429 & 0.0 & 3.4 & 86.4 & 31{,}345 & 1.5 & 93.1 \\
\bottomrule
\end{tabular}
\caption{The tick events classified. Each event takes the first class it satisfies: price rule (the last quote within 60 minutes on the same UTC day is live and has a midpoint above 0.96 or below 0.04), activation (within a day of the asset's first appearance, or before its first live quote or first trade), operator batch (50 or more tick changes within 60 seconds in one `event', the metadata's grouping of related markets), residual. `Mid' is the median midpoint of the last quote within 60 minutes before the event, live or not, on the side of the market above and below 0.5; `near threshold' is the share of those midpoints between 0.96 and 0.97 or between 0.03 and 0.04; `paired' the share of events whose market's other token changed tick within a minute. Panel B tests activation: if an event is the venue switching a new market on, it should come before the asset's first live quote. Panel C is a logit on all adoptions and a five-percent sample of non-adopting asset-days, with the day-before update rate, trading, placeholder status, price extremeness, spread at the floor, age and neg-risk as covariates. Panel D: a one-sided book is a lone bid at or above 0.96 with no ask, or a lone ask at or below 0.04 with no bid; closing and game start times are the venue's metadata.}
\label{tab:rollout}
\end{table}
 
\paragraph{The events against the price rule.} Table~\ref{tab:rollout} classifies the
1{,}551{,}795 token-level changes of a tick from 0.01 to 0.001 on the archive's clean days; because
the market's other token changes tick within a minute in \SI{99.97}{\percent} of cases, they are
774{,}517 changes at the level of the market. For \SI{92.8}{\percent} of the events the asset's last
quote within the preceding hour was live and had a midpoint above 0.96 or below 0.04, the rule the
venue's documentation stated, and the two sides of the market are equally represented, 719{,}850
events on the high side and 719{,}720 on the low. The classes are assigned in order and overlap:
\SI{71.9}{\percent} of the price-rule events also meet the activation criteria, most of them because
the asset had no archived trade before the event (\SI{50.1}{\percent} of all price-rule events; a
book can reach the edge of the range on quotes alone). The class shares therefore reflect the order of
assignment, and the argument rests on the quotes. The midpoint at the event is not near the
thresholds. Its median is 0.995 on the high side and 0.005 on the low, its fifth percentile on the
high side 0.965 and its ninety-fifth percentile on the low side 0.035, and only
\SI{13.4}{\percent} of events sit within 0.01 of a threshold. In \SI{76.7}{\percent} of the
price-rule events the last quote was a one-sided book, a lone bid at or above 0.96 with no ask or a
lone ask at or below 0.04 with no bid.

The events are timed by the market's end. They fall a median of two hours before the market's
closing time in the venue's metadata (quartiles of half an hour and four and a half hours), and
\SI{86.5}{\percent} fall after the close or less than a day before it (\SI{8.2}{\percent} after
it); on the 1{,}098{,}926 events on sports markets with a recorded kick-off,
they fall a median 1.6~hours after the game began, \SI{80.7}{\percent} after it. The activation
class, 71{,}795 events, does not look like activation either: for \SI{94}{\percent} of its events
the asset's first live quote came before the event, a median 776~minutes before, and its events
fall a median 206~minutes before the close. The residual, \SI{2.6}{\percent}, is the same, a median
204~minutes before the close and \SI{93.1}{\percent} of its sports events after kick-off. Both
classes sit at median midpoints of 0.90 and 0.25 (activation) and 0.935 and 0.06 (residual) on the
two sides, and their timing is consistent with the same events, whose extreme quote the hour-long
look-back did not catch. The events follow the rule the venue's documentation stated
\citep{polymarket-docs-market-channel}. The finer tick arrives when the book's price reaches the
edge of the range, which is typically late in a market's life (on sports markets, late in or after
the game) and within hours of the close, when the book holds little but bids for the likely winner
at 0.99 or asks for the likely loser at 0.01.

On the day before adoption the median midpoint was 0.500 and only \SI{1.1}{\percent} of
midpoints lay beyond the thresholds. A day-before state therefore does not capture the
contemporaneous price-rule trigger: the market may still be a placeholder or a live contest,
while on the adoption day the game progresses, the price reaches the edge and the tick changes. The adoption
hazard, fitted on the day-before state over asset-days still on the 0.01 tick, with the sampled
non-adopters weighted by twenty, has an area under the ROC curve of 0.747. Its coefficient on an
extreme midpoint the day before is negative ($-2.06$): a book already at the edge has usually
changed tick on that same day, and the few that have not are the ones the rule has yet to catch. It
predicts a daily hazard of 0.17 for a placeholder book and 0.35 for a first active day (a day-old asset with a thousand updates and a trade), a ratio of
2.00; a venue switching new markets on would put the first active day's hazard near one.

\begin{table}[htbp]
\centering\scriptsize\setlength{\tabcolsep}{3.5pt}
\begin{tabular}{lrrrrrr}
\toprule
specification & cohorts & treated & pre mean & pre $p$ & post mean & ratio \\
\midrule
baseline, all & 117 & 241{,}401 & 0.597 & $<0.003$ & $-1.472$ & 0.229 \\
baseline, traded on $g-1$ & 117 & 241{,}686 & 0.712 & $<0.003$ & $-1.320$ & 0.267 \\
baseline, stable activity & 117 & 18{,}476 & 0.224 & $<0.003$ & $-1.061$ & 0.346 \\
screened, all & 81 & 210{,}697 & 0.591 & $<0.003$ & $-1.539$ & 0.215 \\
screened, traded on $g-1$ & 81 & 203{,}358 & 0.734 & $<0.003$ & $-1.382$ & 0.251 \\
screened, stable activity & 81 & 15{,}659 & 0.242 & $<0.003$ & $-1.154$ & 0.316 \\
reweighted, all & 81 & 210{,}697 & 0.422 & $<0.003$ & $-1.346$ & 0.260 \\
reweighted, traded on $g-1$ & 81 & 203{,}358 & 0.441 & $<0.003$ & $-1.245$ & 0.288 \\
reweighted, stable activity & 81 & 15{,}659 & $-0.068$ & 0.037 & $-1.017$ & 0.362 \\
\bottomrule
\end{tabular}
\caption{Group-time effects of tick adoption on the log of the day's mean quoted spread, with not-yet-treated controls, aggregated by event time $e\in[-7,7]$ and weighted by cohort size; day $-1$ is the base. The pre-period mean is over $e\le-2$ and its $p$-value is the joint pre-trend test from a bootstrap over cohorts (300 draws); the spread ratio is $\exp$ of the post-period mean. The `all' sample is a fixed one-in-five sample of assets. `Baseline' reports the three unweighted samples; `screened' removes the archive's degraded days, and with them the cohorts whose adoption or base day is degraded; `reweighted' reweights the control units by the odds $p/(1-p)$ of the fitted daily adoption hazard of Table~\ref{tab:rollout}, averaged over each asset's panel days and capped at 20, on the `screened' sample (median weight 0.26, 90th percentile 0.51).}
\label{tab:eventstudy}
\end{table}
 
\begin{figure}[htbp]\centering
\includegraphics[width=0.82\linewidth]{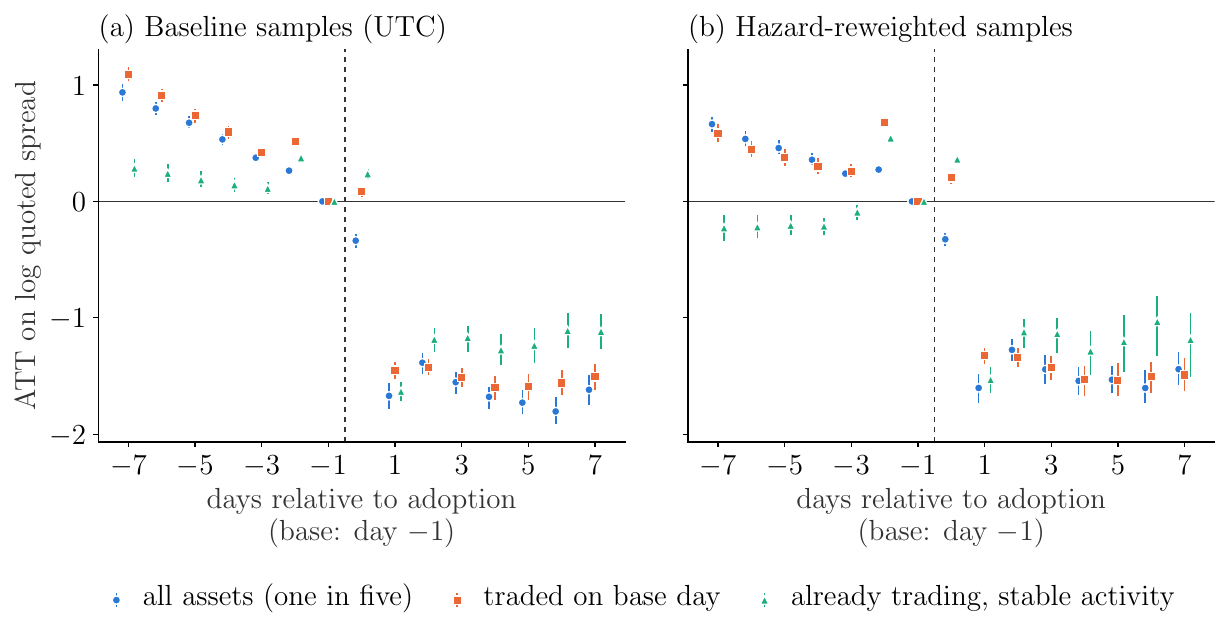}
\caption{Group-time effects of tick adoption on the log of the day's mean quoted spread, aggregated
by event time, with not-yet-treated controls. Panel~(a): the three baseline samples;
panel~(b): the same sample definitions reweighted by the adoption
hazard, without the degraded days. Day $-1$ is the base. Bars are \SI{95}{\percent}
cohort-bootstrap intervals. The path is what a spread does in the days around a market's decision,
not the effect of a tick.}
\label{fig:eventstudy}
\end{figure}

\paragraph{What the event study measures.} Table~\ref{tab:eventstudy} and
Figure~\ref{fig:eventstudy} compare the baseline, health-screened and hazard-reweighted
group-time specifications, all timed in UTC. In every one of them the
joint pre-trend test rejects and the spread falls by more than a log point between the adoption day
and the day after. In all but the reweighted stable-activity sample the pre-period effects are
positive and fall toward the base day until $e=-3$; at $e=-2$ they rise in the traded and
stable-activity samples and, slightly, in the reweighted sample of all adopters. In the
baseline all-asset specification the pre-period mean is 0.597 and the post-period mean
$-1.472$, a spread 0.229 times its base, on 241{,}401 treated units in 117 cohorts. Dropping the
archive's degraded days, and with them the cohorts whose adoption day or base day is degraded,
changes little. Weighting the controls by the odds of their fitted adoption hazard, the weighting
that targets an average effect on the treated, lowers the pre-period mean to 0.422 on all adopters
and leaves the post-period fall at $-1.346$; on the already-trading, stable-activity sample the
pre-period coefficients change sign along the window and their mean, $-0.068$, is not a smaller
pre-trend, while the post-period fall is $-1.017$. Given what the events are, this is the expected
shape. The spread of a market in play narrows as the game approaches its decision and collapses
when the book becomes one bid at 0.99, whose quoted spread is one cent by construction. The
estimator compares those markets with markets that will be decided later, whose spreads are still
those of a live contest, and reweighting the controls cannot remove a pre-trend that is the game
itself. The treated units at later event times are the survivors: with \SI{86.5}{\percent} of
events after the close or less than a day before it, an adopter still quoted at $e=+7$ is a
selected one. We therefore report the event study as a description of the last days of a market's
life and make no claim about the tick from it.

\begin{table}[htbp]
\centering\scriptsize\setlength{\tabcolsep}{3.5pt}
\begin{tabular}{lrrrrrrr}
\toprule
outcome & TWFE & market s.e. & event s.e. & treated & control & paired & perm. $p$ \\
\midrule
\multicolumn{8}{l}{\emph{finer: 0.01 to 0.0025: 40 treated, 363 controls, nominal MDE 0.37; descriptive only}} \\
log spread & $-0.106$ & 0.082 & 0.114 & $-1.010$ & $-0.549$ & $-0.461$ & 0.0175 \\
spread in the old tick & 5.639 & 0.653 & 1.062 & $-0.750$ & $-1.000$ & 0.250 & 0.2769 \\
share of quotes at the tick & $-0.292$ & 0.030 & 0.021 & $-0.013$ & 0.222 & $-0.235$ & 0.0015 \\
log size at best bid & $-2.520$ & 0.150 & 0.210 & $-0.990$ & 1.500 & $-2.490$ & 0.0005 \\
\multicolumn{8}{l}{\emph{coarser: 0.001 to 0.0025: 29 treated, 15 controls, nominal MDE 0.95; descriptive only}} \\
log spread & $-0.044$ & 0.264 & 0.225 & $-0.361$ & $-0.102$ & $-0.258$ & 0.0475 \\
spread in the old tick & 0.882 & 11.205 & 9.804 & $-2.500$ & $-3.500$ & 1.000 & 0.5307 \\
share of quotes at the tick & $-0.104$ & 0.033 & 0.029 & 0.000 & 0.037 & $-0.037$ & 0.0010 \\
log size at best bid & 0.184 & 0.465 & 0.517 & 1.841 & 0.595 & 1.246 & 0.0350 \\
\bottomrule
\end{tabular}
\caption{The 2 July change of World Cup markets to a 0.0025 tick, in twelve-hour bins from 30 June 12:00 to 3 July 24:00 UTC. Controls are the venue's other sports markets of the moneyline, spreads and totals types that were quoted on 1 July, whose game starts between half a day after the change and the end of the window, whose tick before the change matched the arm's and whose pre-period median spread was below 0.5, computed from at least two asset-bins. This spread screen can admit one-sided books near resolution. The outcomes are the log of the half-day median spread, that spread in old ticks, the share of quotes within one old tick and the log of the mean displayed size at the best bid. TWFE is the two-way fixed-effects (market and bin) coefficient on treated times post with standard errors clustered by market and by venue event; the paired columns are the median over markets of the post-period median less the pre-period median, for treated and control markets, and their difference; perm. $p$ is an exploratory two-sided market-label permutation tail share over 2{,}000 draws, with a plus-one correction. World Cup status was not randomised, so this is not causal randomisation inference. The nominal MDE assumes independent markets and 80\% power at the 5\% level. Both arms are descriptive: their treated markets share only six and three events, respectively. The text reports whole-event bootstrap sensitivity.}
\label{tab:worldcup}
\end{table}
 
\paragraph{The World Cup change.} The finer arm, 40 World Cup markets moved from a 0.01 to a 0.0025
tick, has 363 control markets, all of them with a median on both sides of the change, and a nominal minimum
detectable effect of 0.37 on the log half-day median spread. The coarser arm, 29 markets moved from
0.001 to 0.0025, has 15, too few for inference. Table~\ref{tab:worldcup} reports both. The treated
markets are World Cup games played
inside the window, so the comparison is with other sports markets of the same types
whose games start between half a day after the change and the end of the window, on the same tick
before the change and passing the pre-period spread screen. Every market of the finer arm has quotes on 1
and 2 July; both days pass the archive-health rule, although on 1 July the on-chain check found
whole markets missing from the archive.

Before the change the treated books' spread sat at one tick of 0.01; after it, their log median
spread was 1.010 lower, the spread measured in the old tick had
fallen by 0.75, the share of quotes within one old tick had not moved ($-0.013$) and displayed size
at the best bid was 0.99 log points lower, each new tick spanning a quarter of an old one. The
control books also narrowed over these bins without a grid change: their log spread fell
by 0.549 over the same bins, their share of quotes within one old tick rose by 0.222 and their
displayed size at the best rose by 1.500. The paired difference-in-differences on log spread is
therefore $-0.461$. A bootstrap that resamples the six treated and 189 control venue events,
retaining each event's contracts, gives an interval of $[-0.928, -0.144]$; six treated groups
remain too few to support precise inference. The market-label permutation tail share is 0.0175
from 2{,}000 draws, with a plus-one correction; it is an exploratory reference rather than a
randomisation test of the venue's nonrandom assignment. The two-way
fixed-effects estimate, which weights each asset-bin observation rather than each market's paired change, is
$-0.106$ with a market-clustered standard error of 0.082, about a tenth of a log point and not
distinguishable from zero; clustering by event increases the standard error to 0.114. On the spread in old ticks it is $+5.64$ (standard error 0.653): in the mean, the treated spreads widened
relative to controls in units of the old tick, by far more than the paired medians show. The controls have a wider upper tail before the change, so their fall in mean spread
can exceed the treated fall even though the typical treated book reaches a finer floor. The
log transformation, the weighting of asset-bins and the use of paired market medians answer
different questions, and
forty markets over two days cannot settle it. In the coarser arm the treated spread fell by 0.361,
a paired difference of $-0.258$ from the fifteen controls with an interval of
$[-0.654, 0.102]$ and a two-way estimate of $-0.044$ with a standard error of 0.264; with three treated venue events and a nominal minimum
detectable effect of 0.95 the arm cannot identify a spread effect, and its displayed size at the best
rose by 1.841 against 0.595 for the controls, the mirror of the finer arm's fall.

\paragraph{What can be claimed.} On this venue in this period the tick is finer where the price is
extreme. The separate World Cup change supplies a comparison over two days, with forty finer-arm
markets sharing six events. Their typical quoted spread remains one to two ticks before and after,
consistent with a binding grid, while displayed size also changes. Against markets in play on the
old grid, the finer arm's spread is narrower by about half a log point on the paired estimate and
by about a tenth, not distinguishable from zero, on the two-way estimate. Controls narrow as well,
consistent with a common game-time pattern. The comparison cannot attribute the changes to the
grid or to game progress. Both arms are descriptive, and their standard errors and permutation tail shares
in Table~\ref{tab:worldcup} are not used for inference. We do not report a tick effect on liquidity
beyond that.

\subsection{Robustness summary}
\label{sec:robust}

Lee--Ready leads the classifier comparison on every one of the twelve on-chain days,
including the two V1-exchange days, where the side is derived. The prevailing quote is taken
strictly before the print, and the lag grid of Section~\ref{sec:rules} shows what a stale quote
costs; the day-level ranges in Table~\ref{tab:cached-benchmark} give the day-to-day dispersion.
The economic sensitivities, including the loss of the aggregate impact sign reversal when
tied prints are excluded, are reported in Sections~\ref{sec:method-signed} and~\ref{sec:economic-results}.
The cutover's placebo ranks are reported
for three statistics and two days, and again on the subset of dates whose whole window contains no
degraded day; the archive's own format seam is put through a paired change of its own (14 to 16 April); the adverse-tail
sensitivity at $k=7$ is given beside the survivors' change; and the depth outcome is cross-checked
against the book snapshots and reported as not comparable across the migration. The tick event
study compares baseline, health-screened and hazard-reweighted specifications in UTC
(Section~\ref{sec:method-tick}). Its reading rests on what the events are, which Table~\ref{tab:rollout} establishes four
ways: the class shares, the quotes at the event, the activation test and the timing against the
close and the kick-off. The World Cup arms state their minimum detectable effect before the
estimate and report market-clustered standard errors, a market bootstrap of the paired estimate,
event-level sensitivity checks and exploratory market-label permutations where the distribution
is not degenerate. Both World Cup arms remain descriptive. The archive-health rule that
decides which days enter is stated in full in Appendix~\ref{app:health}.
  
\end{document}